# Binary Linear Locally Repairable Codes

Pengfei Huang, *Student Member, IEEE,* Eitan Yaakobi, *Member, IEEE,*
Hironori Uchikawa, *Member, IEEE,* and Paul H. Siegel, *Fellow, IEEE*

**Abstract**

Locally repairable codes (LRCs) are a class of codes designed for the local correction of erasures. They have received considerable attention in recent years due to their applications in distributed storage. Most existing results on LRCs do not explicitly take into consideration the field size $q$, i.e., the size of the code alphabet. In particular, for the binary case, only a few results are known.

In this work, we present an upper bound on the minimum distance $d$ of linear LRCs with availability, based on the work of Cadambe and Mazumdar. The bound takes into account the code length $n$, dimension $k$, locality $r$, availability $t$, and field size $q$. Then, we study binary linear LRCs in three aspects. First, we focus on analyzing the locality of some classical codes, i.e., cyclic codes and Reed-Muller codes, and their modified versions, which are obtained by applying the operations of extend, shorten, expurgate, augment, and lengthen. Next, we construct LRCs using phantom parity-check symbols and multi-level tensor product structure, respectively. Compared to other previous constructions of binary LRCs with fixed locality or minimum distance, our construction is much more flexible in terms of code parameters, and gives various families of high-rate LRCs, some of which are shown to be optimal with respect to their minimum distance. Finally, availability of LRCs is studied. We investigate the locality and availability properties of several classes of one-step majority-logic decodable codes, including cyclic simplex codes, cyclic difference-set codes, and 4-cycle free regular low-density parity-check (LDPC) codes. We also show the construction of a long LRC with availability from a short one-step majority-logic decodable code.

## I. Introduction

Distributed and cloud storage systems today are required to tolerate the failure or unavailability of some of the nodes in the system. The simplest and most commonly used way to accomplish this task is replication, where every

P. Huang and P.H. Siegel are with the Department of Electrical and Computer Engineering and the Center for Memory Recording Research, University of California, San Diego, La Jolla, CA 92093, U.S.A. (e-mail: {pehuang, psiegel}@ucsd.edu).

E. Yaakobi is with the Department of Computer Science, Technion – Israel Institute of Technology, Haifa 32000, Israel (e-mail: yaakobi@cs.technion.ac.il).

H. Uchikawa is with the Toshiba Corporation, Japan (e-mail: hironori.uchikawa@toshiba.co.jp).

The work of H. Uchikawa was done while he was with the Center for Memory Recording Research, University of California, San Diego, La Jolla, CA 92093, U.S.A.

This research was supported in part by the Lady David Foundation, a Viterbi Research Fellowship, and an Israel Science Foundation (ISF) Grant No. 1624/14. Part of the results in the paper were presented at the IEEE Information Theory Workshop, Jerusalem, Israel, Apr 26 – May 01, 2015 (reference [10]), and the IEEE International Symposium on Information Theory, HongKong, China, June 14 – 19, 2015 (reference [11]).

node is replicated several times, usually three. This solution has clear advantages due to its simplicity and fast recovery from node failures. However, it entails a large storage overhead which becomes costly in large storage systems.

In order to achieve better storage efficiency, erasure codes, e.g., Reed-Solomon codes, are deployed. Reed-Solomon and more generally maximum distance separable (MDS) codes are attractive since they tolerate the maximum number of node failures for a given redundancy. However, they suffer from a very slow recovery process, in the case of a single node failure, which is the most common failure scenario. Hence, an important objective in the design of erasure codes is to ensure fast recovery while efficiently supporting a large number of node failures. There are several metrics in the literature to quantify the efficiency of rebuilding. Three of the most popular consider the number of communicated bits in the network, the number of read bits, and the number of accessed nodes. In this work, we study codes with respect to the last metric.

Locally repairable codes (LRCs) are a class of codes in which a failure of a single node can be recovered by accessing at most $r$ other nodes, where $r$ is a predetermined value [6], [18], [20]. For a length-$n$ code with dimension $k$, it is said that the code has *all-symbol locality* $r$ if every symbol is recoverable from a set of at most $r$ symbols. If the code is systematic and only its information symbols have this property then the code has *information locality* $r$. LRCs are well studied in the literature and many works have considered code constructions and bounds for such codes. In [6], an upper bound, which can be seen as a modified version of the Singleton bound, was given on the minimum distance of LRCs. More specifically, if an $[n,k,d]_q$ linear code has information locality $r$, then

$$d \leqslant n - k - \left\lceil \frac{k}{r} \right\rceil + 2. \tag{1}$$

In [20], it was proved that bound (1) also holds for non-linear codes with all-symbol locality. Code constructions which achieve bound (1) were given in [5], [8], [9], [26], [28], [32], [34]. However, for some cases, bound (1) is not tight, so several improvements were proposed in [22], [30], [35]. Recently, a new upper bound on the dimension $k$ of LRCs was presented in [4]. This bound takes into account the code length, minimum distance, locality, and field size, and it is applicable to both non-linear and linear codes. Namely, if an $(n,M,d)_q$ code has all-symbol locality $r$, then

$$k \leqslant \min_{x \in \mathbb{Z}^+} \left\{ xr + k_{opt}^{(q)}(n - x(r+1), d) \right\}, \tag{2}$$

where $M$ denotes the codebook size, $k = \log_q M$, $\mathbb{Z}^+$ is the set of all positive integers, and $k_{opt}^{(q)}(n',d')$ is the largest possible dimension of a length-$n'$ code with minimum distance $d'$ and a given field size $q$. There also exist some constructions of LRCs over small fields, e.g., binary field, in [7], [10], [11], [29], [33], [40].

In addition to symbol locality, another important property of LRCs is their symbol availability, meaning the number of disjoint sets of symbols that can be used to recover a given symbol. High availability is a particularly attractive property for so-called *hot data* in a distributed storage network. More precisely, a code $\mathcal{C}$ has *all-symbol locality $r$ and availability $t$* if every code symbol can be recovered from $t$ disjoint repair sets of other symbols, each set of size at most $r$ symbols. We refer to such a code as an $(r,t)_a$-LRC. If the code is systematic and these properties apply only to its information symbols, then the code has *information locality $r$ and availability $t$*, and it is referred to as an $(r,t)_i$-LRC.

Several recent works have considered codes with both locality and availability properties. In [36], it was shown that the minimum distance $d$ of an $[n,k,d]_q$ linear $(r,t)_i$-LRC satisfies the upper bound

$$d \leqslant n-k-\left\lceil \frac{t(k-1)+1}{t(r-1)+1} \right\rceil +2. \tag{3}$$

In [23], it was proved that bound (3) is also applicable to $(n,M,d)_q$ non-linear $(r,t)_i$-LRCs. In the same paper, it was also shown that if each repair set in a linear $(r,t)_i$-LRC contains only one parity symbol, then the minimum distance $d$ of the code satisfies the following upper bound

$$d \leqslant n-k-\left\lceil \frac{kt}{r} \right\rceil +t+1, \tag{4}$$

and codes achieving bound (4) were constructed using MDS codes and Gabidulin codes. For $(r,t)_a$-LRCs with parameters $(n,M,d)_q$, it was shown in [31] that $d$ satisfies

$$d \leqslant n-\sum_{i=0}^{t} \left\lfloor \frac{k-1}{r^i} \right\rfloor. \tag{5}$$

There are several constructions of LRCs with availability. In [32], two constructions of $(r,2)_a$-LRCs were proposed. One relies on the combinatorial concept of orthogonal partitions, and the other one is based on product codes. In [19], a class of $(r,t)_a$-LRCs was constructed from partial geometries. A family of systematic fountain codes having information locality and strong probabilistic guarantees on availability was introduced in [2]. More recently, in [7], [40], constructions based on the simplex code were proposed. In [37], a family of LRCs with arbitrary availability was constructed, and it outperforms the direct product codes with respect to the information rate.

In this paper, we study bounds and constructions for linear LRCs over a fixed field size; in particular, we focus on binary linear LRCs. We first develop field size dependent upper bounds that incorporate the availability $t$, based on the work by Cadambe and Mazumdar [4]. For constructions, we make contributions in the following three aspects.

We investigate the locality of two classes of classical codes, i.e., cyclic codes and Reed-Muller codes. We observe that the locality of a cyclic code is determined by the minimum distance of its dual code, and show that this result also holds for Reed-Muller codes. We also discuss the locality of a code obtained when we extend, shorten, expurgate, augment, and lengthen an LRC.

Tensor product codes, first proposed by Wolf in [38], are a family of codes defined by a parity-check matrix that is the tensor product of the parity-check matrices of two constituent codes. Later, they were generalized in [12]. As shown in [39], the encoding steps of tensor product codes involve using *phantom* syndrome symbols, which only appear in the encoding procedure and will disappear in the final codewords. Motivated by these ideas, we give three constructions (**Constructions A**, **B**, and **C**) of LRCs that leverage phantom parity-check symbols. These constructions are effective for LRCs with small minimum distance. To obtain LRCs with higher minimum distance, we present another construction (**Construction D**) based on multi-level tensor product structure. All our constructions are flexible and generate a variety of high-rate LRCs with different localities. Some of these codes are proved to have an optimal minimum distance.

One-step majority-logic decodable codes were first formally studied by Massey [15], [17]. Historically, these codes were introduced for low-complexity error correction. Every symbol of such codes has several disjoint repair sets, and is decoded according to the majority of the values given by all of its repair sets. In this work, we make the connection between one-step majority-logic decodable codes and LRCs with availability. We also demonstrate how a long $(r,t)_a$-LRC can be constructed from a short one-step majority-logic decodable code using multi-level tensor product structure.

The rest of the paper is organized as follows. In Section II, we formally define the problem and present field size $q$ dependent bounds on the minimum distance $d$ and the dimension $k$ of $[n,k,d]_q$ linear $(r,t)_i$-LRCs. In Section III, we investigate the locality of several classical codes and their modified versions obtained using standard code operations. In Section IV, we construct various families of $(r,1)_i$-LRCs and $(r,1)_a$-LRCs using phantom parity-check symbols and multi-level tensor product structure. In Section V, we review several families of one-step majority-logic decodable codes, and identify the locality and availability of these codes. Section VI concludes the paper.

## II. Definitions and Bounds

We begin with several basic definitions and notational conventions. We use the notation $[n]$ to define the set $\{1,\ldots,n\}$. For a length-$n$ vector $\boldsymbol{v}$ and a set $\mathcal{I} \subseteq [n]$, the vector $\boldsymbol{v}_{\mathcal{I}}$ denotes the restriction of the vector $\boldsymbol{v}$ to coordinates in the set $\mathcal{I}$. A linear code $\mathcal{C}$ over $\mathbb{F}_q$ of length $n$, dimension $k$, and minimum distance $d$ will be denoted by $[n,k,d]_q$, and its generator matrix is $G = (g_1,\ldots,g_n)$, where $g_i \in \mathbb{F}_q^k$ is a column vector for $i \in [n]$. We define $k_{\mathcal{I}}(\mathcal{C}) = \log_q |\{\boldsymbol{c}_{\mathcal{I}} : \boldsymbol{c} \in \mathcal{C}\}|$, and, for simplicity, we write $k_{\mathcal{I}}$ instead of $k_{\mathcal{I}}(\mathcal{C})$ when $\mathcal{C}$ is clear from the context. The dual code of a linear code $\mathcal{C}$ will be denoted by $\mathcal{C}^{\perp}$.

We follow the conventional definitions of linear LRCs with availability, as established in [23], [31], [36].

**Definition 1.** *The $i$th code symbol of an $[n,k,d]_q$ linear code $\mathcal{C}$ is said to have locality $r$ and availability $t$ if there exist $t$ pairwise disjoint repair sets $\mathcal{R}_i^1,\ldots,\mathcal{R}_i^t \subseteq [n]\backslash\{i\}$, such that 1) $|\mathcal{R}_i^j| \leqslant r$, for $1 \leqslant j \leqslant t$, and 2) for each repair set $\mathcal{R}_i^j$, $1 \leqslant j \leqslant t$, $g_i$ is a linear combination of the columns $g_u$, $u \in \mathcal{R}_i^j$.*

**Definition 2.** *Let $\mathcal{C}$ be an $[n,k,d]_q$ linear code. A set $\mathcal{I} \subseteq [n]$ is said to be an information set if $|\mathcal{I}| = k_{\mathcal{I}} = k$.*
*1) The code $\mathcal{C}$ is said to have all-symbol locality $r$ and availability $t$ if every code symbol has locality $r$ and availability $t$. We refer to $\mathcal{C}$ as a linear $(r,t)_a$-LRC.*
*2) The code $\mathcal{C}$ is said to have information locality $r$ and availability $t$ if there is an information set $\mathcal{I}$ such that, for any $i \in \mathcal{I}$, the $i$th code symbol has locality $r$ and availability $t$. We refer to $\mathcal{C}$ as a linear $(r,t)_i$-LRC.*

Note that when $t=1$, Definition 2 reduces to the definition of linear LRCs. It is straightforward to verify that the minimum distance $d$ of a linear $(r,t)_a$-LRC satisfies $d \geqslant t+1$. We now present upper bounds on the minimum distance and the dimension of linear $(r,t)_i$-LRCs, based on the framework established in [4]. The following lemma and theorem are extensions of Lemma 1 and Theorem 1 from [4], respectively, and for the completeness of the results in the paper we provide their detailed proofs in Appendix A and Appendix B.

Let $r$ and $x$ be two positive integers and $\boldsymbol{y} = (y_1, \ldots, y_x) \in ([t])^x$ be a vector of $x$ positive integers. We define the integers $A(r, x, \boldsymbol{y})$ and $B(r, x, \boldsymbol{y})$ as follows,

$$A(r, x, \boldsymbol{y}) = \sum_{j=1}^{x} (r-1) y_j + x,$$

$$B(r, x, \boldsymbol{y}) = \sum_{j=1}^{x} r y_j + x.$$

**Lemma 3.** *Let $\mathcal{C}$ be an $[n, k, d]_q$ linear $(r, t)_i$-LRC. Assume that $x \in \mathbb{Z}^+$ and $\boldsymbol{y} = (y_1, \ldots, y_x) \in ([t])^x$ satisfy $1 \leqslant x \leqslant \lceil \frac{k-1}{(r-1)t+1} \rceil$ and $A(r, x, \boldsymbol{y}) < k$. Then, there exists a set $\mathcal{I} \subseteq [n]$ such that $|\mathcal{I}| = B(r, x, \boldsymbol{y})$ and $k_{\mathcal{I}}(\mathcal{C}) \leqslant A(r, x, \boldsymbol{y})$.*

Now, let $d^{(q)}_{\ell-opt}[n, k]$ denote the largest possible minimum distance of an $[n, k, d]_q$ linear code, and let $k^{(q)}_{\ell-opt}[n, d]$ denote the largest possible dimension of such a code. Applying Lemma 3, we get the following upper bounds on $d$ and $k$ for $[n, k, d]_q$ linear $(r, t)_i$-LRCs.

**Theorem 4.** *For any $[n, k, d]_q$ linear $(r, t)_i$-LRC, the minimum distance $d$ satisfies*

$$d \leqslant \min_{\substack{1 \leqslant x \leqslant \lceil \frac{k-1}{(r-1)t+1} \rceil,\ x \in \mathbb{Z}^+, \\ \boldsymbol{y} \in ([t])^x, \\ A(r,x,\boldsymbol{y}) < k}} \left\{ d^{(q)}_{\ell-opt}[n - B(r, x, \boldsymbol{y}), k - A(r, x, \boldsymbol{y})] \right\}, \tag{6}$$

*and the dimension satisfies*

$$k \leqslant \min_{\substack{1 \leqslant x \leqslant \lceil \frac{k-1}{(r-1)t+1} \rceil,\ x \in \mathbb{Z}^+, \\ \boldsymbol{y} \in ([t])^x, \\ A(r,x,\boldsymbol{y}) < k}} \left\{ A(r, x, \boldsymbol{y}) + k^{(q)}_{\ell-opt}[n - B(r, x, \boldsymbol{y}), d] \right\}. \tag{7}$$

**Remark 1** Since a linear $(r, t)_a$-LRC is also a linear $(r, t)_i$-LRC, bounds (6) and (7) hold for linear $(r, t)_a$-LRCs as well.

## III. Locality of Binary Classical Codes and Their Modified Versions

In this section, we study the all-symbol locality of classical codes and their modified versions. We also investigate their optimality, in the sense of the following definition.

**Definition 5.** *An $[n, k, d]_q$ linear code $\mathcal{C}$ with all-symbol locality $r$ is said to be $d$-optimal if there does not exist an $[n, k, d+1]_q$ code with all-symbol locality $r$. Similarly, it is called $k$-optimal if there does not exist an $[n, k+1, d]_q$ code with all-symbol locality $r$. Finally, it is called $r$-optimal if there does not exist an $[n, k, d]_q$ code with all-symbol locality $r-1$.*

**Example 1.** Consider the binary simplex code $\mathcal{C}$ with parameters $[2^m - 1, m, 2^{m-1}]$. It was proved in [4] that this code has all-symbol locality $r = 2$ and it is $r$-optimal for these given parameters. Since this code satisfies the Plotkin bound, it is $d$-optimal and $k$-optimal as well.

In the remainder of this section, we consider only codes with all-symbol locality, and thus when saying that a code has locality $r$ we refer to all-symbol locality.

### A. Locality of Classical Codes

In this subsection, we study two classes of binary classical codes, namely, cyclic codes and Reed-Muller codes.

First, we give our main result for cyclic codes, and also present several examples. We start with a simple observation about the locality of code symbols. Even though it has been mentioned before, see e.g., [6], [21], [22], we state it here as a remark for completeness.

**Remark 2** For a binary linear code $\mathcal{C}$, if its $i$th coordinate, $i \in [n]$, belongs to the support of a codeword in $\mathcal{C}^{\perp}$ with weight $r+1$, then the $i$th code symbol has locality $r$.

The next lemma is an immediate consequence of the preceding remark.

**Lemma 6.** *Let $\mathcal{C}$ be an $[n,k,d]$ cyclic binary linear code, and let $d^{\perp}$ be the minimum distance of its dual code $\mathcal{C}^{\perp}$. Then, the code $\mathcal{C}$ has locality $d^{\perp}-1$.*

*Proof:* The dual code $\mathcal{C}^{\perp}$ has a codeword of weight $d^{\perp}$. Since $\mathcal{C}$ is a cyclic linear code, its dual code $\mathcal{C}^{\perp}$ is also a cyclic linear code. Thus every $i \in [n]$ belongs to the support of some codeword of weight $d^{\perp}$ in $\mathcal{C}^{\perp}$. From Remark 2, every coordinate has locality $d^{\perp}-1$. Thus, the code $\mathcal{C}$ has locality $r = d^{\perp}-1$. ■

Next, we give several examples to illustrate how the locality of specific codes can be determined from Lemma 6 and then study their optimality.

**Example 2.** Let $\mathcal{C}$ be the $[n=2^m-1, k=2^m-1-m, d=3]$ cyclic binary Hamming code. Its dual code is the $[2^m-1, m, 2^{m-1}]$ cyclic binary simplex code. Therefore, the Hamming code has locality $r = 2^{m-1}-1$. Since it is a perfect code, it is both $d$-optimal and $k$-optimal. In order to show $r$-optimality, let us assume on the contrary that there exists an $[n,k,d]$ code with locality $\hat{r} = 2^{m-1}-2$. According to bound (2) for $x=1$, we have that

$$\begin{aligned} k \leqslant & x\hat{r} + k_{opt}^{(2)}(n - x(\hat{r}+1), d) = 2^{m-1} - 2 + k_{opt}^{(2)}(2^{m-1}, 3) \\ \overset{(a)}{<} & 2^{m-1} - 2 + 2^{m-1} - (m-1) = 2^m - m - 1, \end{aligned}$$

where step $(a)$ is from the Hamming bound. Thus, we get a contradiction to the value of $k$. We also get from Lemma 6 that the simplex code has locality 2. This gives an alternative proof to the one given in [4] in case the code is cyclic.

**Example 3.** Here we consider the $[23,12,7]$ cyclic binary Golay code $\mathcal{C}$. Its dual code $\mathcal{C}^{\perp}$ is the $[23,11,8]$ cyclic binary code. Hence, we conclude that $\mathcal{C}$ has locality $r=7$ and the dual code $\mathcal{C}^{\perp}$ has locality $r^{\perp}=6$. The code $\mathcal{C}$ is both $d$-optimal and $k$-optimal since it is a perfect code. $\mathcal{C}^{\perp}$ is $d$-optimal due to the Hamming bound, and $k$-optimal according to the online table [27] for $k_{\ell-opt}^{(2)}[23,8]$. The $r$-optimality of these two codes is proved in a similar way to the optimality proof in Example 2.

**Example 4.** Let $\mathcal{C}$ be the cyclic double-error-correcting binary primitive BCH (DBCH) code with parameters $[2^m-1, 2^m-1-2m, 5]$ where $m \geqslant 4$. Its dual code $\mathcal{C}^{\perp}$ has parameters $[2^m-1, 2m, 2^{m-1}-2^{\lfloor m/2 \rfloor}]$ [15]. Therefore, we conclude that $\mathcal{C}$ has locality $r = 2^{m-1}-2^{\lfloor m/2 \rfloor}-1$, and $\mathcal{C}^{\perp}$ has locality $r^{\perp}=4$. We utilize bound (2) and the online table from [27] to check the $d$-optimality, $k$-optimality, and $r$-optimality of the DBCH codes and their dual

TABLE I

PARAMETERS OF DBCH CODES AND THEIR DUAL CODES.

| $\mathcal{C}$ | $n$ | $k$ | $d$ | $r$ | $d$-opt | $k$-opt | $r$-opt |
|---|---|---|---|---|---|---|---|
| $m=4$ | 15 | 7 | 5 | 3 | ✓ | ✓ | ✓ |
| $m=5$ | 31 | 21 | 5 | 11 | ✓ | ✓ | ? |
| $m=6$ | 63 | 51 | 5 | 23 | ✓ | ✓ | ? |
| $m=7$ | 127 | 113 | 5 | 55 | ✓ | ✓ | ? |
| $m=8$ | 255 | 239 | 5 | 111 | ✓ | ✓ | ? |
| $\mathcal{C}^\perp$ | $n^\perp$ | $k^\perp$ | $d^\perp$ | $r^\perp$ | $d$-opt | $k$-opt | $r$-opt |
| $m=4$ | 15 | 8 | 4 | 4 | ✓ | ? | ? |
| $m=5$ | 31 | 10 | 12 | 4 | ✓ | ✓ | ? |
| $m=6$ | 63 | 12 | 24 | 4 | ? | ? | ? |
| $m=7$ | 127 | 14 | 56 | 4 | ✓ | ? | ? |
| $m=8$ | 255 | 16 | 112 | 4 | ? | ? | ? |

codes. The results are summarized in Table I (where ✓ indicates that we could prove optimality while ? means that we could not).

Reed-Muller (RM) codes form another important class of codes. They are simple to construct and rich in structural properties. This motivates us to study their locality. Recall that a $\mu$th-order binary RM code $\mathcal{RM}(\mu, m)$ has code length $n = 2^m$, dimension $k = \sum_{i=0}^{\mu} \binom{m}{i}$, and minimum distance $d = 2^{m-\mu}$.

In [24], two classes of codes with locality 2 and 3 were constructed based on the non-binary RM codes of first and second orders. Here, we focus on the binary RM codes of any order, and determine their locality as follows.

**Lemma 7.** *The $\mu$th-order binary RM code $\mathcal{RM}(\mu, m)$ has locality* $r = d^\perp - 1 = 2^{\mu+1} - 1$.

*Proof:* It is known that the dual code of $\mathcal{RM}(\mu, m)$ is $\mathcal{RM}(m-\mu-1, m)$, and the minimum weight codewords of an RM code generate all of its codewords [15]. Therefore, every coordinate $i$, $i \in [n]$, belongs to the support of a certain minimum weight codeword of $\mathcal{RM}(m-\mu-1, m)$. To see that, assume on the contrary that there exists a coordinate $j$, $j \in [n]$, in which all the minimum weight codewords of $\mathcal{RM}(m-\mu-1, m)$ have value 0. Thus, any linear combinations of the minimum weight codewords cannot produce the all-ones codeword $\mathbf{1}$, which is a valid codeword. Thus, we get a contradiction, which implies that $\mathcal{RM}(\mu, m)$ has locality $r = d^\perp - 1 = 2^{\mu+1} - 1$. ∎

Finally, we mention that a $\mu$th-order cyclic binary RM code $\mathcal{C}$ is a $[2^m - 1, \sum_{i=0}^{\mu} \binom{m}{i}, 2^{m-\mu} - 1]$ punctured binary RM code, represented in a cyclic form [15]. Its dual code $\mathcal{C}^\perp$ is also cyclic and is a $[2^m - 1, \sum_{i=\mu+1}^{m} \binom{m}{i} - 1, 2^{\mu+1}]$ binary code. From Lemma 6, $\mathcal{C}$ has locality $r = 2^{\mu+1} - 1$, and $\mathcal{C}^\perp$ has locality $r^\perp = 2^{m-\mu} - 2$.

## *B. Locality of Modified Classical Codes*

In this subsection, we show how to find the locality of codes which are obtained by applying the standard code operations of extending, shortening, expurgating, augmenting, and lengthening to existing LRCs. For a binary vector $\boldsymbol{c}$, let $\overline{\boldsymbol{c}}$ represent the complement vector of $\boldsymbol{c}$. For a binary code $\mathcal{C}$, define $\overline{\mathcal{C}} = \{\overline{\boldsymbol{c}} \,:\, \boldsymbol{c} \in \mathcal{C}\}$.

*1) Extend Operation:* The extended code of an $[n,k,d]$ binary code $\mathcal{C}$ is an $[n+1,k,d_{ext}]$ code $\mathcal{C}_{ext}$ with an overall parity bit added to each codeword,

$$\mathcal{C}_{ext} = \left\{ (c_1,\ldots,c_n,c_{n+1}) : (c_1,\ldots,c_n) \in \mathcal{C}, c_{n+1} = \sum_{i=1}^{n} c_i \right\},$$

where $d_{ext} = d+1$ for odd $d$ and $d_{ext} = d$ for even $d$. We use the notation $\mathcal{C}_{ext}^{\perp}$ to denote the dual code of $\mathcal{C}_{ext}$.

**Lemma 8.** *Let $\mathcal{C}$ be an $[n,k,d]$ binary code with locality $r$. If the maximum Hamming weight of codewords in $\mathcal{C}^{\perp}$ is $n-r$, then the extended code $\mathcal{C}_{ext}$ has locality $r_{ext} = r$.*

*Proof:* For every $i \in [n]$, there exists a set $R_i$ of size at most $r$ such that the $i$th symbol is recoverable from the set $R_i$. Thus, we only need to prove this property for the $(n+1)$st symbol. Since the maximum weight of codewords in $\mathcal{C}^{\perp}$ is $n-r$, there exists a codeword $\boldsymbol{c} \in \mathcal{C}^{\perp}$ such that $w_H(\boldsymbol{c}) = n-r$. Note also that the vectors $(\boldsymbol{c},0)$ and $\mathbf{1}$ are codewords in $\mathcal{C}_{ext}^{\perp}$. Therefore the vector $\boldsymbol{c}' = (\boldsymbol{c},0)+\mathbf{1}$ is a codeword in $\mathcal{C}_{ext}^{\perp}$ and its Hamming weight is $r+1$. Hence, from Remark 2, we get that the $(n+1)$st symbol can also be recovered by a set of $r$ other symbols. ■

We have the following corollary for cyclic binary linear codes, for which we have already seen that $r = d^{\perp}-1$.

**Corollary 9.** *Let $\mathcal{C}$ be an $[n,k,d]$ cyclic binary code and let $d^{\perp}$ be the minimum distance of its dual code. If the maximum Hamming weight of codewords in $\mathcal{C}^{\perp}$ is $n+1-d^{\perp}$, then the extended code $\mathcal{C}_{ext}$ has locality $r_{ext} = d^{\perp} - 1$.*

**Example 5.** Let $\mathcal{C}$ be the $[2^m-1, 2^m-1-m, 3]$ cyclic binary Hamming code. Its extended code $\mathcal{C}_{ext}$ has parameters $[2^m, 2^m-1-m, 4]$. The dual code $\mathcal{C}^{\perp}$ is the simplex code, whose nonzero codewords have constant Hamming weight $2^{m-1}$. Hence, the condition from Corollary 9 holds and we conclude that the extended Hamming code $\mathcal{C}_{ext}$ has locality $r_{ext} = d^{\perp}-1 = 2^{m-1}-1$. $\mathcal{C}_{ext}$ is both $d$-optimal and $k$-optimal according to the Hamming bound. To show that it is also $r$-optimal, let us assume on the contrary that there exists a $[2^m, 2^m-1-m, 4]$ binary code with locality $\hat{r} = 2^{m-1}-2$. According to bound (2) for $x=1$, we have

$$\begin{aligned} k_{ext} &\leqslant 2^{m-1} - 2 + k_{opt}^{(2)}(2^{m-1}+1,4) \overset{(a)}{=} 2^{m-1} - 2 + k_{opt}^{(2)}(2^{m-1},3) \\ &\overset{(b)}{<} 2^{m-1} - 2 + 2^{m-1} - (m-1) = 2^m - m - 1. \end{aligned}$$

Thus, we get a contradiction to the value of $k_{ext}$. In the above proof, step $(a)$ follows from the property that $A(n,2s-1) = A(n+1,2s)$, where $A(n,d)$ denotes the largest number of codewords $M$ in any binary code $(n,M,d)$ [16]. Step $(b)$ follows from the Hamming bound.

Next, we determine the locality of the dual of the extension of a cyclic code.

**Lemma 10.** *Let $\mathcal{C}$ be an $[n,k,d]$ cyclic binary code with odd minimum distance $d$. Then, the code $\mathcal{C}_{ext}^{\perp}$ has locality $r_{ext}^{\perp} = d$.*

*Proof:* Since $d$ is odd, each codeword with weight $d$ in $\mathcal{C}$ generates a parity-check bit 1. Since $\mathcal{C}$ is cyclic, for any $i \in [n]$, $i$ belongs to the support of some codeword $(\boldsymbol{c},1) \in \mathcal{C}_{ext}$, where $\boldsymbol{c}$ has weight $d$. Moreover, the support of $(\boldsymbol{c},1)$ also contains coordinate $n+1$. Thus, from Remark 2, every symbol of $\mathcal{C}_{ext}^{\perp}$ has locality $d$. ■

**Example 6.** Let $\mathcal{C}$ be the $[n = 2^m - 1, k = 2^m - 1 - m, d = 3]$ cyclic binary Hamming code. Correspondingly, $\mathcal{C}_{ext}^{\perp}$ is the biorthogonal code $[n_{ext}^{\perp} = 2^m, k_{ext}^{\perp} = m + 1, d_{ext}^{\perp} = 2^{m-1}]$ [13]. From Lemma 10, $\mathcal{C}_{ext}^{\perp}$ has locality $r_{ext}^{\perp} = d = 3$. $\mathcal{C}_{ext}^{\perp}$ is both $d$-optimal and $k$-optimal according to the Plotkin bound. To show that $\mathcal{C}_{ext}^{\perp}$ is $r$-optimal, we utilize bound (2) with $x = 1$, and have the following constraint on the dimension of the code,

$$\begin{aligned} k_{ext}^{\perp} = m + 1 &\leqslant r_{ext}^{\perp} + k_{opt}^{(2)}(2^m - (r_{ext}^{\perp} + 1), 2^{m-1}) \\ &\overset{(a)}{\leqslant} r_{ext}^{\perp} + \log_2 \frac{2 \cdot 2^{m-1}}{2 \cdot 2^{m-1} - 2^m + (r_{ext}^{\perp} + 1)} \\ &= r_{ext}^{\perp} + m - \log_2(r_{ext}^{\perp} + 1), \end{aligned}$$

where step $(a)$ is from the Plotkin bound. Therefore, we obtain

$$r_{ext}^{\perp} \geqslant \log_2(r_{ext}^{\perp} + 1) + 1.$$

Thus, we have $r_{ext}^{\perp} \geqslant 3$. Therefore, the code is $r$-optimal.

*2) Shorten Operation:* For an $[n, k, d]$ binary code $\mathcal{C}$, its shortened code $\mathcal{C}_s$ of $\mathcal{C}$ is the set of all codewords in $\mathcal{C}$ that are 0 in a fixed position with that position deleted. Let the last one of the coordinates of $\mathcal{C}$ be the position deleted, then the shortened code $\mathcal{C}_s$ is

$$\mathcal{C}_s = \{(c_1, \dots, c_{n-1}) : (c_1, \dots, c_{n-1}, 0) \in \mathcal{C}\}.$$

We assume here that there is a codeword $\boldsymbol{c} \in \mathcal{C}$ such that $c_n = 1$. Otherwise, we will remove another coordinate satisfying this condition. The code $\mathcal{C}_s$ has parameters $[n - 1, k - 1, d_s \geqslant d]$ and its dual code is denoted by $\mathcal{C}_s^{\perp}$.

**Lemma 11.** *Let $\mathcal{C}$ be an $[n, k, d]$ binary code with locality $r \geqslant 2$. The shortened code $\mathcal{C}_s$ has locality $r$ or $r - 1$.*

*Proof:* Since $\mathcal{C}$ has locality $r$, for all $i \in [n - 1]$, the $i$th code symbol has a repair set $R_i$ with respect to $\mathcal{C}$ of size at most $r$. If $n \notin R_i$ then this symbol has the same repair set also with respect to the shortened code $\mathcal{C}_s$. Otherwise, note that if $\boldsymbol{c} \in \mathcal{C}_s$ then $(\boldsymbol{c}, 0) \in \mathcal{C}$, so we conclude that the $i$th symbol is recoverable also from the set $R_i \setminus \{n\}$. ■

The following is an immediate consequence of Lemma 11 for cyclic binary codes.

**Corollary 12.** *Let $\mathcal{C}$ be an $[n, k, d]$ cyclic binary code whose dual code has minimum distance $d^{\perp} \geqslant 3$. Then, the code $\mathcal{C}_s$ has locality either $d^{\perp} - 2$ or $d^{\perp} - 1$.*

The next example shows that the shortened code can in fact have locality $r - 1$.

**Example 7.** Let $\mathcal{C}$ be the $[2^m - 1, 2^m - 1 - m, 3]$ cyclic binary Hamming code. Its shortened code $\mathcal{C}_s$ is a $[2^m - 2, 2^m - 2 - m, 3]$ code and from Corollary 12 it has locality $d^{\perp} - 2$ or $d^{\perp} - 1$, where $d^{\perp} = 2^{m-1}$. We show that it has locality $d^{\perp} - 2$. According to the proof of Lemma 11, it is enough to show that for every $i \in [n - 1]$, the $i$th code symbol has a repair set $R_i$ of size $2^{m-1} - 1$ which contains the $n$th coordinate. Or, according to Remark 2, it is enough to show that there exists a codeword $\boldsymbol{c} \in \mathcal{C}^{\perp}$ such that $c_i = c_n = 1$ and $w_H(\boldsymbol{c}) = 2^{m-1}$. We can omit the last requirement on the weight since all nonzero codewords in $\mathcal{C}^{\perp}$ have the same weight $2^{m-1}$. Let $\boldsymbol{c}_1, \boldsymbol{c}_2 \in \mathcal{C}^{\perp}$ be two codewords such that $c_{1,i} = c_{2,n} = 1$. If $c_{1,n} = 1$ or $c_{2,i} = 1$ then we are done. Otherwise, the codeword

TABLE II

LOCALITY OF BINARY CLASSICAL CODES AND THEIR MODIFIED VERSIONS.

| $\mathcal{C}$ | $n$ | $k$ | $d$ | $r$ | *d-opt* | *k-opt* | *r-opt* |
|---|---|---|---|---|---|---|---|
| Hamming code | $2^m-1$ | $2^m-1-m$ | 3 | $2^{m-1}-1$ | ✓ | ✓ | ✓ |
| Simplex code | $2^m-1$ | $m$ | $2^{m-1}$ | 2 | ✓ | ✓ | ✓[a] |
| Golay code | 23 | 12 | 7 | 7 | ✓ | ✓ | ✓ |
| Dual of Golay code | 23 | 11 | 8 | 6 | ✓ | ✓ | ✓ |
| DBCH code $(m \geqslant 4)$ | $2^m-1$ | $2^m-1-2m$ | 5 | $2^{m-1}-2^{\lfloor m/2 \rfloor}-1$ | Table I | Table I | Table I |
| Dual of DBCH code $(m \geqslant 4)$ | $2^m-1$ | $2m$ | $2^{m-1}-2^{\lfloor m/2 \rfloor}$ | 4 | Table I | Table I | Table I |
| Extended Hamming code | $2^m$ | $2^m-1-m$ | 4 | $2^{m-1}-1$ | ✓ | ✓ | ✓ |
| Extended Golay code | 24 | 12 | 8 | 7 | ✓ | ✓ | ✓ |
| Extended DBCH code $(m \geqslant 4)$ | $2^m$ | $2^m-1-2m$ | 6 | $2^{m-1}-2^{\lfloor m/2 \rfloor}-1$ | ✓ | ? | ? |
| Extended TBCH code $(m \geqslant 5)$ | $2^m$ | $2^m-1-3m$ | 8 | $2^{m-1}-2^{\lfloor m/2+1 \rfloor}-1$ | ✓ | ? | ? |
| Biorthogonal code | $2^m$ | $m+1$ | $2^{m-1}$ | 3 | ✓ | ✓ | ✓ |
| Expurgated Hamming code | $2^m-1$ | $2^m-2-m$ | 4 | $2^{m-1}-2$ | ✓ | ✓ | ✓ |
| Expurgated DBCH code $(m \geqslant 4)$ | $2^m-1$ | $2^m-2-2m$ | 6 | $2^{m-1}-2^{\lfloor m/2 \rfloor}-2$ | ✓ | ? | ? |
| Expurgated TBCH code $(m \geqslant 5)$ | $2^m-1$ | $2^m-2-3m$ | 8 | $2^{m-1}-2^{\lfloor m/2+1 \rfloor}-2$ | ✓ | ? | ? |
| Augmented simplex code | $2^m-1$ | $m+1$ | $2^{m-1}-1$ | 3 | ✓ | ✓ | ✓ |
| Shortened Hamming code | $2^m-2$ | $2^m-2-m$ | 3 | $2^{m-1}-2$ | ✓ | ✓ | ✓ |
| Shortened simplex code | $2^m-2$ | $m-1$ | $2^{m-1}$ | 1 | ✓ | ✓ | ✓ |
| $\mathcal{RM}(\mu,m)$ | $2^m$ | $\sum_{i=0}^{\mu}\binom{m}{i}$ | $2^{m-\mu}$ | $2^{\mu+1}-1$ | ? | ? | ? |
| Cyclic $\mathcal{RM}(\mu,m)$ | $2^m-1$ | $\sum_{i=0}^{\mu}\binom{m}{i}$ | $2^{m-\mu}-1$ | $2^{\mu+1}-1$ | ? | ? | ? |
| Dual of cyclic $\mathcal{RM}(\mu,m)$ | $2^m-1$ | $\sum_{i=\mu+1}^{m}\binom{m}{i}-1$ | $2^{\mu+1}$ | $2^{m-\mu}-2$ | ? | ? | ? |

[a] $r$-optimality is proved in [4].

$\boldsymbol{c}_1+\boldsymbol{c}_2$ satisfies this property. The $d$-optimality, $k$-optimality, and $r$-optimality of $\mathcal{C}_s$ are proved in a similar way to the previous examples.

*3) Expurgate, Augment, and Lengthen Operations:* For an $[n,k,d]$ binary code $\mathcal{C}$ having at least one odd weight codeword, the expurgated code $\mathcal{C}_{exp}$ is a subcode of $\mathcal{C}$ which contains only the codewords of even weight. That is,

$$\mathcal{C}_{exp}=\{\boldsymbol{c}:\boldsymbol{c}\in\mathcal{C}, w_H(\boldsymbol{c}) \text{ is even }\}.$$

$\mathcal{C}_{exp}$ is an $[n,k-1,w_e]$ code, where $w_e$ denotes the minimum even weight of nonzero codewords in $\mathcal{C}$. We denote by $\mathcal{C}_{exp}^{\perp}$ the dual code of $\mathcal{C}_{exp}$ and note that $\mathcal{C}_{exp}^{\perp}=\mathcal{C}^{\perp}\cup\overline{\mathcal{C}^{\perp}}$.

For an $[n,k,d]$ binary code $\mathcal{C}$ which does not contain the all-ones codeword $\mathbf{1}$, the augmented code $\mathcal{C}_a$ is the code $\mathcal{C}\cup\overline{\mathcal{C}}$ with parameters $[n,k+1,\min\{d,n-w_{\max}\}]$, where $w_{\max}$ denotes the maximum weight of codewords in $\mathcal{C}$. We use the notation $\mathcal{C}_a^{\perp}$ to denote the dual code of $\mathcal{C}_a$.

According to these definitions, if the code $\mathcal{C}$ is cyclic then the expurgated and augmented codes of $\mathcal{C}$ are cyclic as well. Hence, for an $[n,k,d]$ cyclic binary code $\mathcal{C}$, we have the following two observations:

a) If $\mathcal{C}$ has an odd weight codeword, then $\mathcal{C}_{exp}$ has locality $r_{exp}=\min\{d^{\perp},n-w_{\max}^{\perp}\}-1$, where $w_{\max}^{\perp}$ is the maximum weight of codewords in $\mathcal{C}^{\perp}$. (Here, we assume $w_{\max}^{\perp}<n-1$, since $w_{\max}^{\perp}=n-1$ is not an interesting case.)

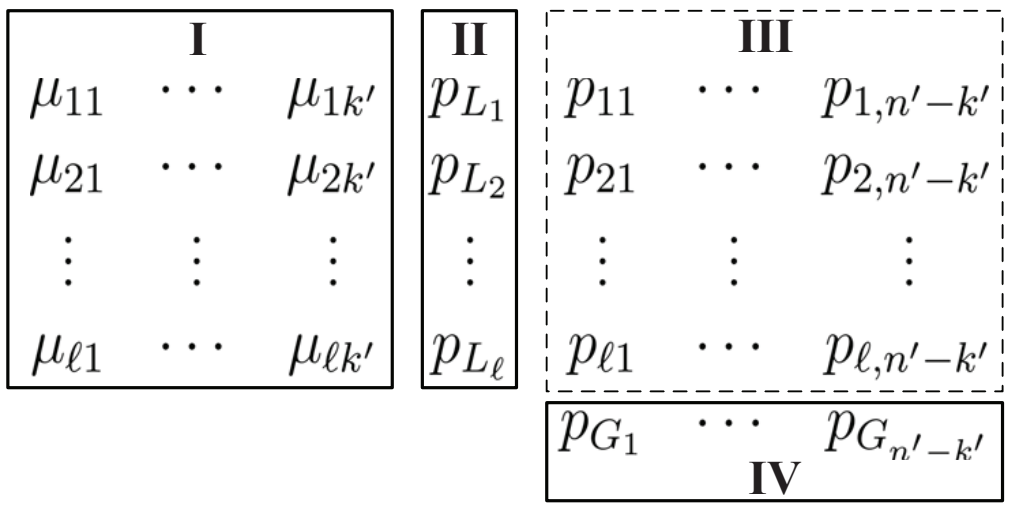

Fig. 1. An $(r, 1)_i$-LRC using *Construction A*. Information symbols are in block **I**, local parity-check symbols are in block **II**, phantom symbols are in block **III**, and global parity-check symbols are in block **IV**.

b) If $\mathcal{C}$ does not contain the all-ones codeword **1**, then $\mathcal{C}_a$ has locality $r_a = w_e^{\perp} - 1$, where $w_e^{\perp}$ is the minimum even weight of nonzero codewords in $\mathcal{C}^{\perp}$.

For an $[n, k, d]$ binary code $\mathcal{C}$ which does not contain the all-ones codeword **1**, the lengthened code $\mathcal{C}_\ell$ is obtained as follows. First, the code $\mathcal{C}$ is augmented to the code $\mathcal{C}_a = \mathcal{C} \cup \overline{\mathcal{C}}$. Then, $\mathcal{C}_a$ is extended. Thus, $\mathcal{C}_\ell = \{(c_1, \ldots, c_n, c_{n+1}) : c_{n+1} = \sum_{i=1}^{n} c_i \text{ and } (c_1, \ldots, c_n) \in \mathcal{C} \cup \overline{\mathcal{C}}\}$. After the lengthen operation, the length and dimension of the code are increased by 1. By leveraging the results from the augment and extend operations, we conclude that if the minimum even weight of nonzero codewords in $\mathcal{C}^{\perp}$ is $w_e^{\perp}$, and the maximum weight of codewords in $\mathcal{C}_a^{\perp}$ is $n + 1 - w_e^{\perp}$, then the lengthened code $\mathcal{C}_\ell$ has locality $r_\ell = w_e^{\perp} - 1$.

Our results on locality of binary classical codes and their modified versions are summarized in Table II, where ✓ means we can prove the optimality of the given codes, whereas ? means we have not verified their optimality.

## IV. Construction of Binary LRCs

In this section, we focus on constructing binary LRCs. We first present constructions of LRCs with small minimum distance (i.e., $d = 3$, 4, and 5) by using phantom parity-check symbols. Then, in order to obtain LRCs with higher minimum distance, we propose another construction which is based on multi-level tensor product structure.

### A. Construction Using Phantom Parity-Check Symbols

We first consider constructing binary linear $(r, 1)_i$-LRCs with minimum distance 3 and 4. The general framework is depicted in Fig. 1. We specify an $[n', k', d']$ systematic binary code as a base code, $\mathcal{C}_{base}$. The following construction produces an $(r, 1)_i$-LRC of length $n = (k' + 1)\ell + n' - k'$, dimension $k = k'\ell$, and information locality $r = k'$.

**Construction A**

**Step 1:** Place an $\ell \times k'$ array of information symbols in block **I**.

**Step 2:** For each row of information symbols, $(\mu_{i1}, \ldots, \mu_{ik'})$, $1 \leqslant i \leqslant \ell$, compute *local* parity-check symbols $p_{L_i} = \sum_{j=1}^{k'} \mu_{ij}$, $1 \leqslant i \leqslant \ell$, and place them in the corresponding row of block **II**.

**Step 3:** Encode each row of information symbols in block **I** using $\mathcal{C}_{base}$, producing parity-check symbols $(p_{i1}, \ldots, p_{i,n'-k'})$, $1 \leqslant i \leqslant \ell$. Place these parity-check symbols in block **III**. (These symbols are referred to as *phantom* symbols because they will not appear in the final codeword.)

**Step 4:** Compute a row of *global* parity-check symbols, $p_{G_j} = \sum_{i=1}^{\ell} p_{ij}$, $1 \leqslant j \leqslant n' - k'$, by summing the rows of phantom symbols in block **III**. Place these symbols in block **IV**.

**Step 5:** The constructed codeword consists of the symbols in blocks **I**, **II**, and **IV**. ■

Note that for $r|k$, Pyramid codes are optimal $(r,1)_i$-LRCs over sufficiently large field size [8]. A Pyramid code is constructed by splitting a parity-check symbol of a systematic MDS code into $k/r$ local parity-check symbols. However, for the binary case, it is hard to find a good binary code first and then conduct the splitting operation. In contrast, we take a different approach. We first design the local parity-check symbols, and then construct the global parity-check symbols.

If $\mathcal{C}_{base}$ has an *information-sum parity-check symbol*, a parity-check symbol which is the sum of all its information symbols, we can simply modify Step 3 of *Construction A* to reduce the code redundancy as follows. After encoding each row of information symbols in block **I**, define the corresponding row of phantom symbols to be the computed parity-check symbols with the information-sum parity-check symbol excluded, and store them in block **III**. Then proceed with the remaining steps in *Construction A*. We refer to this modified construction as **Construction A′**. It is easy to verify that the resulting code is an $(r,1)_i$-LRC with length $n = (k'+1)\ell + n' - k' - 1$, dimension $k = k'\ell$, and information locality $r = k'$.

Now, if we use a $\mathcal{C}_{base}$ with minimum distance 3, we have a lower bound on the minimum distance of the constructed LRC, as stated in the following lemma.

**Lemma 13.** *If $\mathcal{C}_{base}$ is an $[n', k', d' = 3]$ code, the $(r,1)_i$-LRC produced by Construction A (or Construction A′, if appropriate) has minimum distance $d \geqslant 3$.*

*Proof:* See Appendix C. ■

Based on Lemma 13, we have the following theorem on the construction of $(r,1)_i$-LRCs with optimal minimum distance $d = 3$.

**Theorem 14.** *Let $\mathcal{C}_{base}$ be an $[n', k', d' = 3]$ binary code with an information-sum parity-check symbol and assume that $d^{(2)}_{\ell-opt}[n', k'] = 3$. The $(r,1)_i$-LRC obtained from Construction A′ has parameters $[n = (k'+1)\ell + n' - k' - 1, k = k'\ell, d = 3]$ and $r = k'$. Its minimum distance $d = 3$ is optimal.*

*Proof:* From *Construction A′*, the length, dimension and locality of the $(r,1)_i$-LRC are determined. From Lemma 13, the minimum distance satisfies $d \geqslant 3$. On the other hand, from bound (6)[1], with $x = \ell - 1$ and $t = 1$, $d \leqslant d^{(2)}_{\ell-opt}[n - (k'+1)(\ell-1), k - k'(\ell-1)] = d^{(2)}_{\ell-opt}[n', k'] = 3$. Therefore, $d = 3$ and it is optimal. ■

We give some examples of $(r,1)_i$-LRCs with $d = 3$. First, let $\mathcal{C}_{base}$ be the $[7,4,3]$ systematic binary Hamming code whose parity-check matrix is

$$H_{[7,4,3]} = \begin{bmatrix} 0 & 1 & 1 & 1 & 1 & 0 & 0 \\ 1 & 1 & 1 & 0 & 0 & 1 & 0 \\ 1 & 1 & 0 & 1 & 0 & 0 & 1 \end{bmatrix}.$$

[1]Note that we use here and henceforth bound (6) instead of bound (2) simply because bound (6) is stated explicitly for the minimum distance and bound (2) is given for the code dimension.

Using *Construction A*, we obtain an $(r,1)_i$-LRC with parameters $[5\ell+3, 4\ell, 3]$ with $r=4$. However, the upper bound on the minimum distance from bound (6) is 4. To construct an $(r,1)_i$-LRC whose minimum distance is optimal with respect to bound (6), we use a $[6,3,3]$ shortened binary Hamming code as the $\mathcal{C}_{base}$ whose parity-check matrix $H_{[6,3,3]}$ is obtained by deleting the first column of $H_{[7,4,3]}$,

$$H_{[6,3,3]} = \begin{bmatrix} 1 & 1 & 1 & 1 & 0 & 0 \\ 1 & 1 & 0 & 0 & 1 & 0 \\ 1 & 0 & 1 & 0 & 0 & 1 \end{bmatrix}.$$

Now, the $\mathcal{C}_{base}$ has an information-sum parity-check symbol and $d^{(2)}_{\ell-opt}[6,3]=3$. From Theorem 14, the $(r,1)_i$-LRC generated by *Construction A′* has parameters $[4\ell+2, 3\ell, 3]$ and $r=3$. Moreover, its minimum distance $d=3$ is optimal.

The above $[6,3,3]$ base code $\mathcal{C}_{base}$ can be generalized as follows. Let $\mathcal{C}$ be a $[2^m-1, 2^m-1-m, 3]$ systematic binary Hamming code with parity-check matrix

$$H = \begin{bmatrix} h_{1,1} & h_{1,2} & \dots & h_{1,2^m-1-m} & 1 & 0 & \dots & 0 \\ h_{2,1} & h_{2,2} & \dots & h_{2,2^m-1-m} & 0 & 1 & \dots & 0 \\ \vdots & \vdots & \ddots & \vdots & \vdots & \vdots & \ddots & \vdots \\ h_{m,1} & h_{m,2} & \dots & h_{m,2^m-1-m} & 0 & 0 & \dots & 1 \end{bmatrix},$$

whose columns range over all the nonzero vectors in $\mathbb{F}_2^m$. The first $2^m-1-m$ coordinates of $\mathcal{C}$ form the systematic information symbols. The parity-check matrix $H_s$ of the shortened binary Hamming code $\mathcal{C}_s$ is obtained by deleting any $i$th column of $H$, if $1 \leqslant i \leqslant 2^m-1-m$ and $h_{1,i}=0$. As a result, $\mathcal{C}_s$ is systematic and has an information-sum parity-check symbol.

**Lemma 15.** *The code $\mathcal{C}_s$ has parameters $[2^{m-1}+m-1, 2^{m-1}-1, 3]$, and its minimum distance is optimal.*

*Proof:* The first row of $H$ has $2^{m-1}$ ones and $2^{m-1}-1$ zeros, since it is a nonzero codeword of the $[2^m-1, m, 2^{m-1}]$ binary simplex code. According to the shortening operation, we delete in total $2^{m-1}-m$ columns from $H$, so the length of $\mathcal{C}_s$ becomes $2^{m-1}+m-1$ and the dimension becomes $2^{m-1}-1$. Since the shortening operation does not decrease the minimum distance, and there always exist three dependent columns in $H_s$ (e.g., $[1,1,0,\dots,0]^T$, $[1,0,0,\dots,0]^T$, and $[0,1,0,\dots,0]^T$), the minimum distance remains 3. Lastly, we have that $d^{(2)}_{\ell-opt}[2^{m-1}+m-1, 2^{m-1}-1]=3$ from the anticode bound [1]. ■

The following example is a direct result of Theorem 14 and Lemma 15.

**Example 8.** Let $\mathcal{C}_{base}$ be the shortened binary Hamming code $\mathcal{C}_s$ in Lemma 15. The $(r,1)_i$-LRC obtained from *Construction A′* has parameters $[2^{m-1}\ell+m-1, (2^{m-1}-1)\ell, 3]$ and $r=2^{m-1}-1$. Its minimum distance is optimal.

Next, we use a code $\mathcal{C}_{base}$ with minimum distance 4, and have the following lemma.

**Lemma 16.** *If $\mathcal{C}_{base}$ is an $[n', k', d'=4]$ code, the $(r,1)_i$-LRC produced by Construction A (or Construction A′, if appropriate) has minimum distance $d \geqslant 4$.*

*Proof:* The proof is similar to the one of Lemma 13. ■

Based on Lemma 16, we have the following two theorems on the construction of $(r,1)_i$-LRCs with optimal minimum distance $d=4$.

**Theorem 17.** *Let $\mathcal{C}_{base}$ be an $[n',k',d'=4]$ binary code with $d^{(2)}_{\ell-opt}[n'+1,k']=4$. The $(r,1)_i$-LRC obtained from Construction A has parameters $[n=(k'+1)\ell+n'-k',k=k'\ell,d=4]$ and $r=k'$. Its minimum distance $d=4$ is optimal.*

*Proof:* The proof is similar to the one of Theorem 14. ■

**Theorem 18.** *Let $\mathcal{C}_{base}$ be an $[n',k',d'=4]$ binary code with an information-sum parity-check symbol and $d^{(2)}_{\ell-opt}[n',k']=4$. The $(r,1)_i$-LRC obtained from Construction A′ has parameters $[n=(k'+1)\ell+n'-k'-1,k=k'\ell,d=4]$ and $r=k'$. Its minimum distance $d=4$ is optimal.*

*Proof:* The proof is similar to the one of Theorem 14. ■

We give examples of $(r,1)_i$-LRCs with $d=4$ using expurgated or extended binary Hamming code as $\mathcal{C}_{base}$. The following lemma gives properties of expurgated and extended binary Hamming codes.

**Lemma 19.** *For $m\geqslant 4$, the $[2^m-1,2^m-2-m,4]$ systematic expurgated binary Hamming code has no information-sum parity-check symbol, and $d^{(2)}_{\ell-opt}[2^m,2^m-2-m]=4$. For $m\geqslant 3$, the $[2^m,2^m-1-m,4]$ systematic extended binary Hamming code has no information-sum parity-check symbol, and $d^{(2)}_{\ell-opt}[2^m+1,2^m-1-m]=4$.*

*Proof:* For the expurgated binary Hamming code, in its dual code, except the all-ones codeword with weight $2^m-1$, there is no codeword with weight larger than $2^{m-1}$. If the expurgated binary Hamming code has an information-sum parity-check symbol, then in its dual code there is a codeword with weight $2^m-1-m$, which is larger than $2^{m-1}$ for $m\geqslant 4$. We have $d^{(2)}_{\ell-opt}[2^m,2^m-2-m]=4$ from the Hamming bound. Similarly, for the extended binary Hamming code, in its dual code, except the all-ones codeword with weight $2^m$, there is no codeword with weight larger than $2^{m-1}$. If the extended binary Hamming code has an information-sum parity-check symbol, then in its dual code there is a codeword with weight $2^m-m$, which is larger than $2^{m-1}$ for $m\geqslant 3$. We have $d^{(2)}_{\ell-opt}[2^m+1,2^m-1-m]=4$ from the Hamming bound. ■

The following example presents $(r,1)_i$-LRCs with $d=4$ from Theorem 17 and Lemma 19.

**Example 9.** Let $\mathcal{C}_{base}$ be the $[2^m-1,2^m-2-m,4]$ expurgated binary Hamming code, where $m\geqslant 4$. The $(r,1)_i$-LRC obtained from *Construction A* has parameters $[(2^m-1-m)\ell+m+1,(2^m-2-m)\ell,4]$ and $r=2^m-2-m$. Its minimum distance 4 is optimal. Similarly, let $\mathcal{C}_{base}$ be the $[2^m,2^m-1-m,4]$ extended binary Hamming code, where $m\geqslant 3$. The $(r,1)_i$-LRC obtained from *Construction A* has parameters $[(2^m-m)\ell+m+1,(2^m-1-m)\ell,4]$ and $r=2^m-1-m$. Its minimum distance 4 is optimal.

Next, we give a construction of $(r,1)_i$-LRCs for $d=5$. Let $\mathcal{C}_{base}$ be an $[n',k',5]$ systematic binary code, and let $\mathcal{C}'_{base}=\{\boldsymbol{c}_{[k'+w]}:\boldsymbol{c}\in\mathcal{C}_{base}\}$, i.e., restrict $\mathcal{C}_{base}$ to $k'$ information coordinates and $w$ parity-check coordinates, where $w$ is chosen properly such that $\mathcal{C}'_{base}$ has minimum distance at least 3. The following new construction is based on two rows of global parity-check symbols as shown in Fig. 2.

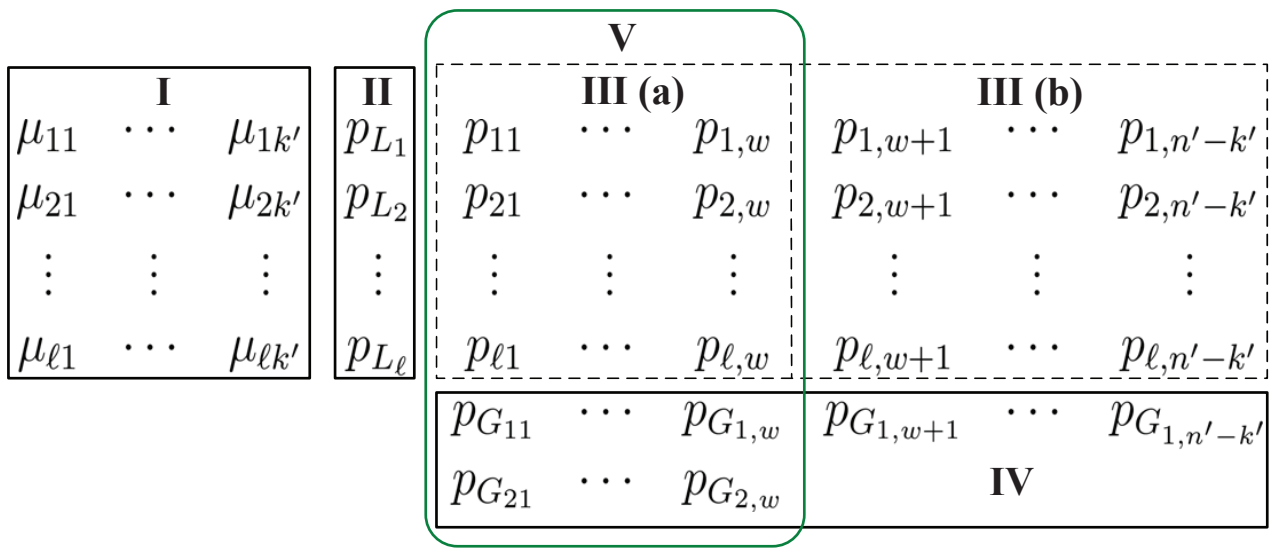

Fig. 2. An $(r,1)_i$-LRC using *Construction B*.

**Construction B**

**Step 1:** Follow Steps 1, 2, and 3 of *Construction A* to get *local* parity-check symbols and *phantom* symbols.

**Step 2:** Divide phantom symbols into two parts: $w$ columns in block **III(a)** and the rest of the columns in block **III(b)**.

**Step 3:** Compute *global* parity-check symbols in block **IV**: 1) Follow Step 4 of *Construction A* to get the first row $(p_{G_{11}}, \cdots, p_{G_{1,n'-k'}})$. 2) Use an $[\ell+2, \ell, 3]$ doubly extended Reed-Solomon code to encode the phantom symbols in block **III(a)** to get the second row $(p_{G_{21}}, \cdots, p_{G_{2,w}})$, by taking each row in block **III(a)** as a symbol in $\mathbb{F}_{2^w}$.

**Step 4:** The constructed codeword consists of the symbols in blocks **I**, **II**, and **IV**. ■

Let $\alpha$ be a primitive element in $\mathbb{F}_{2^w}$, and $\ell \leqslant 2^w - 1$. Then, the parity-check matrix for the doubly extended Reed-Solomon code in *Construction B* is

$$H = \begin{bmatrix} 1 & 1 & 1 & \cdots & 1 & 1 & 0 \\ 1 & \alpha & \alpha^2 & \cdots & \alpha^{\ell-1} & 0 & 1 \end{bmatrix}.$$

Note that an alternative to the doubly extended Reed-Solomon code is an EVENODD code [3].

**Theorem 20.** *The $(r,1)_i$-LRC obtained from Construction B has parameters $[n = (k'+1)\ell + n' - k' + w, k = k'\ell, d \geqslant 5]$, where $\ell \leqslant 2^w - 1$. It has information locality $r = k'$.*

*Proof:* See Appendix D. ■

**Example 10.** Let $\mathcal{C}_{base}$ be the $[2^m - 1, 2^m - 1 - 2m, 5]$ binary BCH code where $m \geqslant 4$. For the case of $m = 4$, exhaustive search shows that we can choose $w$ to be 4. For $\ell \leqslant 15$, the $(r,1)_i$-LRC from *Construction B* has parameters $[n = 8\ell + 12, k = 7\ell, d = 5]$ and $r = 7$. An upper bound on $d$ from bound (6) is 8. For the case of $m = 5$, exhaustive search shows that we can choose $w$ to be 6. For $\ell \leqslant 63$, the $(r,1)_i$-LRC from *Construction B* has parameters $[n = 22\ell + 16, k = 21\ell, d = 5]$ and $r = 21$. An upper bound on $d$ from bound (6) is 8.

We finish this subsection with a construction of $(r,1)_a$-LRCs with minimum distance 4 by using phantom parity-check symbols. We start with an $[n', k', d']$ systematic binary code as a base code, $\mathcal{C}_{base}$. For simplicity, we assume that $k' \geqslant n' - k'$. We use Fig. 1 to illustrate our construction of $(r,1)_a$-LRCs as follows.

**Construction C**

**Step 1:** Place an $\ell \times k'$ array of symbols in block **I**. The first $\ell - 1$ rows are all information symbols. The last row has $2k' - n'$ information symbols (i.e., $\mu_{\ell 1}, \ldots, \mu_{\ell,2k'-n'}$) and $n' - k'$ zero symbols (i.e., $\mu_{\ell,2k'-n'+1} = 0, \ldots, \mu_{\ell,k'} = 0$).
**Step 2:** Encode each row of symbols in block **I** using $\mathcal{C}_{base}$, producing parity-check symbols $(p_{i1}, \ldots, p_{i,n'-k'})$, $1 \leqslant i \leqslant \ell$. Place these parity-check symbols in block **III** as *phantom* symbols.
**Step 3:** Compute a row of *global* parity-check symbols, $p_{G_j} = \sum_{i=1}^{\ell} p_{ij}$, $1 \leqslant j \leqslant n' - k'$, by summing the rows of phantom symbols in block **III**. Place these symbols in block **IV**.
**Step 4:** Let $\mu_{\ell,2k'-n'+j} = p_{G_j}$, $1 \leqslant j \leqslant n' - k'$. For each row of symbols, $(\mu_{i1}, \ldots, \mu_{ik'})$, $1 \leqslant i \leqslant \ell$, compute *local* parity-check symbols $p_{L_i} = \sum_{j=1}^{k'} \mu_{ij}$, $1 \leqslant i \leqslant \ell$, and place them in the corresponding row of block **II**.
**Step 5:** The constructed codeword consists of the symbols in blocks **I** and **II**. ■

The resulting code is an $(r, 1)_a$-LRC of code length $n = (k' + 1)\ell$, dimension $k = k'\ell - (n' - k')$, and all-symbol locality $r = k'$.

We give the following theorem on the construction of $(r, 1)_a$-LRCs with optimal minimum distance 4.

**Theorem 21.** *Let $\mathcal{C}_{base}$ be an $[n', k', d' = 4]$ systematic binary code with $k' \geqslant n' - k'$ and $d^{(2)}_{\ell-opt}[k' + 1, 2k' - n'] \leqslant 4$. The $(r, 1)_a$-LRC obtained from Construction C has parameters $[n = (k' + 1)\ell, k = k'\ell - (n' - k'), d = 4]$ and all-symbol locality $r = k'$. Its minimum distance $d = 4$ is optimal.*

*Proof:* From *Construction C*, the length, dimension and locality of the $(r, 1)_a$-LRC are determined. On one hand, the minimum distance $d \geqslant 4$ since the $(r, 1)_a$-LRC can correct any 3 erasures (The proof is similar to the one of Lemma 13, so we omit it here). On the other hand, from bound (6), with $x = \ell - 1$ and $t = 1$, $d \leqslant d^{(2)}_{\ell-opt}[k' + 1, 2k' - n'] \leqslant 4$. ■

We give the following example of $(r, 1)_a$-LRCs with $d = 4$.

**Example 11.** Let $\mathcal{C}_{base}$ be the $[n' = 2^m - 1, k' = 2^m - 2 - m, d' = 4]$ expurgated binary Hamming code, where $m \geqslant 4$. Since $d^{(2)}_{\ell-opt}[k' + 1, 2k' - n'] = d^{(2)}_{\ell-opt}[2^m - m - 1, 2^m - 2m - 3] \leqslant 4$ due to the Hamming bound, from Theorem 21, the $(r, 1)_a$-LRC obtained from *Construction* C has parameters $[(2^m - 1 - m)\ell, (2^m - 2 - m)\ell - 1 - m, 4]$ and all-symbol locality $r = 2^m - 2 - m$. Its minimum distance 4 is optimal. Similarly, let $\mathcal{C}_{base}$ be the $[2^m, 2^m - 1 - m, 4]$ extended binary Hamming code, where $m \geqslant 3$. The $(r, 1)_a$-LRC obtained from *Construction* C has parameters $[(2^m - m)\ell, (2^m - 1 - m)\ell - 1 - m, 4]$ and all-symbol locality $r = 2^m - 1 - m$. Its minimum distance 4 is optimal.

The binary LRCs constructed in this subsection are summarized in Table III.

### *B. Construction Using Multi-level Tensor Product Structure*

In the previous subsection, we presented constructions of binary LRCs with small minimum distance (i.e., $d = 3$, 4, and 5) based on phantom parity-check symbols. Here, we propose a new construction by using the multi-level tensor product structure [12], leading to $(r, 1)_a$-LRCs with higher minimum distance.

We start by presenting the tensor product operation of two matrices $H^{'}$ and $H^{''}$. Let $H^{'}$ be the parity-check matrix of a binary code with length $n'$ and dimension $n' - v$. $H^{'}$ can be considered as a $v$ (row) by $n'$ (column)

TABLE III

CONSTRUCTED BINARY LRCS IN SECTION IV-A

| $(r,1)_i$-LRCs | $n$ | $k$ | $d$ | $r$ |
|---|---|---|---|---|
| Example 8 | $2^{m-1}\ell+m-1$ | $(2^{m-1}-1)\ell$ | 3 | $2^{m-1}-1$ |
| Example 9 | $(2^m-1-m)\ell+m+1$ | $(2^m-2-m)\ell$ | 4 | $2^m-2-m$ |
| Example 9 | $(2^m-m)\ell+m+1$ | $(2^m-1-m)\ell$ | 4 | $2^m-1-m$ |
| Example 10 | $8\ell+12$ ($\ell\leqslant 15$) | $7\ell$ | 5 | 7 |
| Example 10 | $22\ell+16$ ($\ell\leqslant 63$) | $21\ell$ | 5 | 21 |
| $(r,1)_a$-LRCs | $n$ | $k$ | $d$ | $r$ |
| Example 11 | $(2^m-1-m)\ell$ | $(2^m-2-m)\ell-1-m$ | 4 | $2^m-2-m$ |
| Example 11 | $(2^m-m)\ell$ | $(2^m-1-m)\ell-1-m$ | 4 | $2^m-1-m$ |

matrix over $\mathbb{F}_2$ or as a 1 (row) by $n'$ (column) matrix of elements from $\mathbb{F}_{2^v}$. Let $H^{'}=[h_1^{'}\ h_2^{'}\ \cdots\ h_{n'}^{'}]$, where $h_j^{'}$, $1\leqslant j\leqslant n'$, are elements of $\mathbb{F}_{2^v}$. Let $H^{''}$ be the parity-check matrix of a code of length $\ell$ and dimension $\ell-\lambda$ over $\mathbb{F}_{2^v}$. We denote $H^{''}$ by

$$H^{''}=\begin{bmatrix} h_{11}^{''} & \cdots & h_{1\ell}^{''} \\ \vdots & \ddots & \vdots \\ h_{\lambda 1}^{''} & \cdots & h_{\lambda\ell}^{''} \end{bmatrix},$$

where $h_{ij}^{''}$, $1\leqslant i\leqslant\lambda$ and $1\leqslant j\leqslant\ell$, are elements of $\mathbb{F}_{2^v}$.

The tensor product of the two matrices $H^{'}$ and $H^{''}$ is defined as

$$H^{''}\bigotimes H^{'}=\begin{bmatrix} h_{11}^{''}H^{'} & \cdots & h_{1\ell}^{''}H^{'} \\ \vdots & \ddots & \vdots \\ h_{\lambda 1}^{''}H^{'} & \cdots & h_{\lambda\ell}^{''}H^{'} \end{bmatrix},$$

where $h_{ij}^{''}H^{'}=[h_{ij}^{''}h_1^{'}\ h_{ij}^{''}h_2^{'}\ \cdots\ h_{ij}^{''}h_{n'}^{'}]$, $1\leqslant i\leqslant\lambda$ and $1\leqslant j\leqslant\ell$, and the products of elements are calculated according to the rules of multiplication for elements over $\mathbb{F}_{2^v}$.

Our construction of $(r,1)_a$-LRCs is based on the multi-level tensor product structure proposed in [12]. Define the matrices $H_i^{'}$ and $H_i^{''}$ $(i=1,2,\ldots,\mu)$ as follows. $H_i^{'}$ is a $v_i\times n'$ matrix over $\mathbb{F}_2$ such that the $(v_1+v_2+\cdots+v_i)\times n'$ matrix

$$B_i=\begin{bmatrix} H_1^{'} \\ H_2^{'} \\ \vdots \\ H_i^{'} \end{bmatrix}$$

is a parity-check matrix of an $[n',n'-v_1-v_2-\cdots-v_i,d_i^{'}]$ binary code. $H_i^{''}$ is a $\lambda_i\times\ell$ matrix over $\mathbb{F}_{2^{v_i}}$, which is a parity-check matrix of an $[\ell,\ell-\lambda_i,\delta_i]_{2^{v_i}}$ code.

We define a $\mu$-level tensor product code as a binary linear code having a parity-check matrix in the form of the following $\mu$-level tensor product structure

$$H = \begin{bmatrix} H_1'' \otimes H_1' \\ H_2'' \otimes H_2' \\ \vdots \\ H_\mu'' \otimes H_\mu' \end{bmatrix}. \tag{8}$$

We denote this code by $\mathcal{C}_{TP}^{\mu}$. Its length is $n = n'\ell$ and the number of parity-check symbols is $n - k = \sum_{i=1}^{\mu} v_i \lambda_i$.

Let us give an example of a 2-level tensor product code $\mathcal{C}_{TP}^{2}$.

**Example 12.** Let $H_1' = [1\ 1\ 1\ 1\ 1\ 1\ 1]$ over $\mathbb{F}_2$, and

$$H_2' = \begin{bmatrix} 0 & 0 & 0 & 1 & 1 & 1 & 1 \\ 0 & 1 & 1 & 0 & 0 & 1 & 1 \\ 1 & 0 & 1 & 0 & 1 & 0 & 1 \end{bmatrix}$$

over $\mathbb{F}_2$. Let $H_1'' = [1\ 1\ 1]$ over $\mathbb{F}_2$ and

$$H_2'' = \begin{bmatrix} 1 & 1 & 0 \\ 1 & 0 & 1 \end{bmatrix}$$

over $\mathbb{F}_8$. Hence, in this construction, we use the following parameters: $n' = 7$, $\ell = 3$, $v_1 = 1$, $v_2 = 3$, $\lambda_1 = 1$, $\lambda_2 = 2$, $\delta_1 = 2$, $\delta_2 = 3$, $d_1' = 2$ and $d_2' = 4$. The binary parity-check matrix $H$ of the 2-level tensor product code $\mathcal{C}_{TP}^{2}$ is

$$H = \begin{bmatrix} H_1'' \otimes H_1' \\ H_2'' \otimes H_2' \end{bmatrix}$$

$$= \begin{bmatrix} 1\,1\,1\,1\,1\,1\,1 & 1\,1\,1\,1\,1\,1\,1 & 1\,1\,1\,1\,1\,1\,1 \\ 0\,0\,0\,1\,1\,1\,1 & 0\,0\,0\,1\,1\,1\,1 & 0\,0\,0\,0\,0\,0\,0 \\ 0\,1\,1\,0\,0\,1\,1 & 0\,1\,1\,0\,0\,1\,1 & 0\,0\,0\,0\,0\,0\,0 \\ 1\,0\,1\,0\,1\,0\,1 & 1\,0\,1\,0\,1\,0\,1 & 0\,0\,0\,0\,0\,0\,0 \\ 0\,0\,0\,1\,1\,1\,1 & 0\,0\,0\,0\,0\,0\,0 & 0\,0\,0\,1\,1\,1\,1 \\ 0\,1\,1\,0\,0\,1\,1 & 0\,0\,0\,0\,0\,0\,0 & 0\,1\,1\,0\,0\,1\,1 \\ 1\,0\,1\,0\,1\,0\,1 & 0\,0\,0\,0\,0\,0\,0 & 1\,0\,1\,0\,1\,0\,1 \end{bmatrix}.$$

The code length is $n = n'\ell = 21$ and the dimension is $k = n - \sum_{i=1}^{2} v_i \lambda_i = 14$. It is possible to verify that every 3 columns of $H$ are linearly independent, but columns 1, 2, 5 and 6 of $H$ are linearly dependent. Therefore, the minimum distance of the code is $d = 4$.

Next, we give the following lemma on the minimum distance of a $\mu$-level tensor product code $\mathcal{C}_{TP}^{\mu}$.

**Lemma 22.** *Assume the following inequalities hold: 1) $d_\mu' \leqslant \delta_1$, and 2) $d_\mu' \leqslant \delta_i d_{i-1}'$, for $i = 2, 3, \ldots, \mu$. Then, the minimum distance $d$ of the $\mu$-level tensor product code $\mathcal{C}_{TP}^{\mu}$ is $d_\mu'$.*

*Proof:* First, we show that $d \leqslant d'_\mu$. For $i = 1, 2, \ldots, \mu$, let $H_i^{'} = [h_1^{'}(i), h_2^{'}(i), \ldots, h_{n'}^{'}(i)]$ over $\mathbb{F}_{2^{v_i}}$, and let $[h_{11}^{''}(i), h_{21}^{''}(i), \ldots, h_{\lambda_i 1}^{''}(i)]^T$ over $\mathbb{F}_{2^{v_i}}$ be the first column of $H_i^{''}$. Since the code with parity-check matrix $B_\mu$ has minimum distance $d'_\mu$, there exist $d'_\mu$ columns of $B_\mu$, say in the set of positions $J = \{b_1, b_2, \ldots, b_{d'_\mu}\}$, which are linearly dependent. That is $\sum_{j \in J} h_j^{'}(i) = 0$, for $i = 1, 2, \ldots, \mu$. Thus, we have $\sum_{j \in J} h_{p1}^{''}(i) h_j^{'}(i) = h_{p1}^{''}(i) \Big( \sum_{j \in J} h_j^{'}(i) \Big) = 0$, for $p = 1, 2, \ldots, \lambda_i$ and $i = 1, 2, \ldots, \mu$. That is, the columns in positions $b_1, b_2, \ldots, b_{d'_\mu}$ of $H$ are linearly dependent.

The inequality $d \geqslant d'_\mu$ is shown in the proof of Theorem 2 in [12]. ■

**Remark 3** Lemma 22 is a modified version of Theorem 2 in [12], which incorrectly states that the minimum distance of a $\mu$-level tensor product code $\mathcal{C}_{TP}^{\mu}$ is the largest integer $d_m$ satisfying the following inequalities: 1) $d_m \leqslant d'_\mu$, 2) $d_m \leqslant \delta_1$, and 3) $d_m \leqslant \delta_i d'_{i-1}$, $i = 2, 3, \ldots, \mu$. If this were true, the 2-level $\mathcal{C}_{TP}^{2}$ code in Example 12, with $\delta_1 = 2$, $\delta_2 = 3$, $d'_1 = 2$ and $d'_2 = 4$, would have minimum distance 2. However, the true minimum distance is 4. Theorem 2 only gives a lower bound on the minimum distance, which we have used in the proof of Lemma 22.

We now present a construction of $(r, 1)_a$-LRCs based on the multi-level tensor product structure.

**Construction D**

**Step 1:** Choose $v_i \times n'$ matrices $H_i^{'}$ over $\mathbb{F}_2$ and $\lambda_i \times \ell$ matrices $H_i^{''}$ over $\mathbb{F}_{2^{v_i}}$, for $i = 1, 2, \ldots, \mu$, which satisfy the following two properties:

1) $H_1^{'} = [1, 1, \cdots, 1]$, i.e., a length-$n'$ all-ones vector, and $H_1^{''} = \mathbf{I}_{\ell \times \ell}$, i.e., an $\ell \times \ell$ identity matrix.

2) The matrices $H_i^{'}$ and $H_i^{''}$ are chosen such that $d'_\mu \leqslant \delta_i d'_{i-1}$, for $i = 2, 3, \cdots, \mu$.

**Step 2:** Generate the parity-check matrix $H$ of the $(r, 1)_a$-LRC according to (8) with the matrices $H_i^{'}$ and $H_i^{''}$, for $i = 1, 2, \ldots, \mu$. ■

**Theorem 23.** *The binary $(r, 1)_a$-LRC from Construction D has length $n = n'\ell$, dimension $k = n'\ell - \sum_{i=1}^{\mu} v_i \lambda_i$, minimum distance $d = d'_\mu$, and all-symbol locality $r = n' - 1$.*

*Proof:* According to *Construction D*, the code length $n = n'\ell$ and dimension $k = n'\ell - \sum_{i=1}^{\mu} v_i \lambda_i$ are determined by the construction of the multi-level tensor product codes. From property 1) in Step 1, the tensor product matrix $H_1^{''} \otimes H_1^{'}$ in $H$ gives all-symbol locality $r = n' - 1$. Since $\delta_1 = \infty$ ($H_1^{''}$ is the identity matrix), $d'_1 = 2$, and $d'_\mu \leqslant \delta_i d'_{i-1}$, we conclude from Lemma 22, that the minimum distance of the constructed $(r, 1)_a$-LRC is $d = d'_\mu$. ■

*Construction D* gives a general method to construct $(r, 1)_a$-LRCs, but not an explicit construction. Next, we give a specific code design.

Let $n' = 2^m - 1$ and $\alpha$ be a primitive element of $\mathbb{F}_{2^m}$. In *Construction D*, for $i = 2, 3, \cdots, \mu$, we choose $H_i^{'} = [\beta^0, \beta^1, \cdots, \beta^{n'-1}]$ where $\beta = \alpha^{2i-3}$. Thus, $B_i$ is the parity-check matrix of an expurgated binary BCH code, so we have $d'_i = 2i$. We also choose $H_i^{''}$ to be the parity-check matrix of an $[\ell, \ell - \lambda_i, \delta_i = \lceil \frac{\mu}{i-1} \rceil]_{2^m}$ code, so we have $d'_\mu = 2\mu \leqslant \delta_i d'_{i-1} = 2(i-1)\lceil \frac{\mu}{i-1} \rceil$. We refer to the $(r, 1)_a$-LRC constructed according to the above design as $\mathcal{C}_{LRC}$, and conclude with the following corollary.

**Corollary 24.** *The $(r,1)_a$-LRC $\mathcal{C}_{LRC}$ has parameters $[(2^m-1)\ell, (2^m-2)\ell - m\sum_{i=2}^{\mu}\lambda_i, 2\mu]$ and all-symbol locality $r = 2^m - 2$.*

In particular, for the construction of the $\mathcal{C}_{LRC}$, in order to minimize the value of $\lambda_i$, we can choose $H_i^{''}$ to be the parity-check matrix of an $[\ell, \ell-\delta_i+1, \delta_i = \lceil \frac{\mu}{i-1} \rceil]_{2^m}$ MDS code, where we require that $\ell \leqslant 2^m+1$ only for the case $\mu > 2$. Thus, the resulting $(r,1)_a$-LRC has parameters $[(2^m-1)\ell, (2^m-2)\ell - m\sum_{i=2}^{\mu}(\lceil \frac{\mu}{i-1} \rceil - 1), 2\mu]$ and all-symbol locality $r = 2^m-2$. We refer to this particular $(r,1)_a$-LRC as $\mathcal{C}_{\mathrm{I}}$. We give some instances of $\mathcal{C}_{\mathrm{I}}$ as follows.

**Example 13.** For $\mu = 2$, $\mathcal{C}_{\mathrm{I}}$ is a $[(2^m-1)\ell, (2^m-2)\ell - m, 4]$ LRC with $r = 2^m-2$. It has an optimal minimum distance with respect to bound (6). For $\mu = 3$ and $\ell \leqslant 2^m+1$, $\mathcal{C}_{\mathrm{I}}$ is a $[(2^m-1)\ell, (2^m-2)\ell - 3m, 6]$ LRC with $r = 2^m-2$. For $\mu = 4$ and $\ell \leqslant 2^m+1$, $\mathcal{C}_{\mathrm{I}}$ is a $[(2^m-1)\ell, (2^m-2)\ell - 5m, 8]$ LRC with $r = 2^m-2$.

In the design of the code $\mathcal{C}_{LRC}$, we can also choose $H_i^{''}$ to be the parity-check matrix of a non-MDS code to remove the length constraint on $\ell$. We illustrate this design with the following example.

**Example 14.** For the $\mathcal{C}_{LRC}$ with $\mu = 3$, we choose $H_2^{''}$ to be the parity-check matrix of an $[\ell = \frac{2^{ms}-1}{2^m-1}, \frac{2^{ms}-1}{2^m-1} - s, 3]_{2^m}$ non-binary Hamming code and $H_3^{''} = [1,1,\cdots,1]$. The resulting $(r,1)_a$-LRC has parameters $[2^{ms}-1, \frac{(2^m-2)(2^{ms}-1)}{2^m-1} - (s+1)m, 6]$ and all-symbol locality $r = 2^m-2$. For the $\mathcal{C}_{LRC}$ with $\mu = 4$, we choose $H_2^{''}$ to be the parity-check matrix of an $[\ell = 2^{2m}+1, 2^{2m}-3, 4]_{2^m}$ non-binary code (see problem 3.44 in [25]), $H_3^{''} = [1,1,\cdots,1]$, and $H_4^{''} = [1,1,\cdots,1]$. The resulting $(r,1)_a$-LRC has parameters $[(2^{2m}+1)(2^m-1), (2^{2m}+1)(2^m-2) - 6m, 8]$ and all-symbol locality $r = 2^m-2$. In general, we can choose the matrix $H_i^{''}$ for $i = 2,3,\ldots,\mu$ to be the parity-check matrix of an $[\ell = 2^{ms}-1, \ell-\lambda_i \geqslant \ell - s(\lceil \frac{\mu}{i-1} \rceil - 1), \delta_i \geqslant \lceil \frac{\mu}{i-1} \rceil]_{2^m}$ non-binary BCH code [16]. The resulting $(r,1)_a$-LRC has parameters $[n = (2^{ms}-1)(2^m-1), k \geqslant (2^{ms}-1)(2^m-2) - ms\sum_{i=2}^{\mu}(\lceil \frac{\mu}{i-1} \rceil - 1), d = 2\mu]$ and all-symbol locality $r = 2^m-2$. We refer to this code as $\mathcal{C}_{\mathrm{I}}'$.

**Remark 4** There exist other choices of the matrices $H_i^{'}$ and $H_i^{''}$ in *Construction D*. For example, we can choose $H_i^{'}$ so that $B_i$ is the parity-check matrix of an extended binary BCH code, and choose $H_i^{''}$ to be the parity-check matrix of an MDS code. Then, the resulting $(r,1)_a$-LRC has parameters $[2^m\ell, (2^m-1)\ell - m\sum_{i=2}^{\mu}(\lceil \frac{\mu}{i-1} \rceil - 1), 2\mu]$ and all-symbol locality $r = 2^m-1$, where we require that $\ell \leqslant 2^m+1$ if $\mu > 2$. We refer to this code as $\mathcal{C}_{\mathrm{II}}$.

The $(r,1)_a$-LRCs constructed in this subsection are summarized in Table IV.

### *C. Comparison to Existing Results*

In this subsection, we summarize our constructions of binary LRCs and compare them with previous results.

Our constructions of binary $(r,1)_i$-LRCs and $(r,1)_a$-LRCs have the following features.

1) They provide LRCs with a wide range of values of minimum distance and locality. This diversity is based on the flexible choices of the base code $\mathcal{C}_{base}$ for *Construction A*, *B*, *C*, and of the matrices $H_i^{'}$ and $H_i^{''}$ for

TABLE IV

CONSTRUCTED BINARY $(r,1)_a$-LRCS IN SECTION IV-B

| *Code* | $n$ | $k$ | $d$ | $r$ |
|---|---|---|---|---|
| $\mathcal{C}_{\mathrm{I}}$ | $(2^m-1)\ell$ | $(2^m-2)\ell - m\sum_{i=2}^{\mu}(\lceil\frac{\mu}{i-1}\rceil-1)$ | $2\mu$ | $2^m-2$ |
| $\mathcal{C}_{\mathrm{I}}(\mu=2)$ | $(2^m-1)\ell$ | $(2^m-2)\ell-m$ | 4 | $2^m-2$ |
| $\mathcal{C}_{\mathrm{I}}(\mu=3)$ | $(2^m-1)\ell$ | $(2^m-2)\ell-3m$ | 6 | $2^m-2$ |
| $\mathcal{C}_{\mathrm{I}}(\mu=4)$ | $(2^m-1)\ell$ | $(2^m-2)\ell-5m$ | 8 | $2^m-2$ |
| $\mathcal{C}_{\mathrm{II}}$ | $2^m\ell$ | $(2^m-1)\ell - m\sum_{i=2}^{\mu}(\lceil\frac{\mu}{i-1}\rceil-1)$ | $2\mu$ | $2^m-1$ |
| $\mathcal{C}_{\mathrm{II}}(\mu=2)$ | $2^m\ell$ | $(2^m-1)\ell-m$ | 4 | $2^m-1$ |
| $\mathcal{C}_{\mathrm{II}}(\mu=3)$ | $2^m\ell$ | $(2^m-1)\ell-3m$ | 6 | $2^m-1$ |
| $\mathcal{C}_{\mathrm{II}}(\mu=4)$ | $2^m\ell$ | $(2^m-1)\ell-5m$ | 8 | $2^m-1$ |
| Example 14 | $2^{ms}-1$ | $\frac{(2^m-2)(2^{ms}-1)}{2^m-1}-(s+1)m$ | 6 | $2^m-2$ |
| Example 14 | $(2^{2m}+1)(2^m-1)$ | $(2^{2m}+1)(2^m-2)-6m$ | 8 | $2^m-2$ |
| $\mathcal{C}_{\mathrm{I}}'$ | $(2^{ms}-1)(2^m-1)$ | $(2^{ms}-1)(2^m-2)-ms\sum_{i=2}^{\mu}(\lceil\frac{\mu}{i-1}\rceil-1)$ | $2\mu$ | $2^m-2$ |

For $\mathcal{C}_{\mathrm{I}}$ and $\mathcal{C}_{\mathrm{II}}$, we require $\ell \leqslant 2^m+1$ when $\mu > 2$.

TABLE V

EXISTING CONSTRUCTIONS OF BINARY $(r,1)_a$-LRCS

| *Code* | $n$ | $k$ | $d$ | $r$ |
|---|---|---|---|---|
| [7] | $2^m-1$ $(2\mid m)$ | $\frac{2}{3}(2^m-1)-m$ | 6 | 2 |
| [7] | $2^m-1$ $(2\mid m)$ | $\frac{2}{3}(2^m-1)-2m$ | 10 | 2 |
| [40] | $2^m+1$ $(2\nmid m)$ | $\frac{2}{3}(2^m+1)-2m$ | 10 | 2 |
| [40] | $(2^r+1)(r+1)$ | $(2^r-1)r$ | 6 | $r$ |
| [29] | $2^m-\binom{s}{2}-1$ $(s\leqslant m)$ | $m$ | $2^{m-1}-\lfloor\frac{s^2}{4}\rfloor$ | 2 |
| [29] | $2^m-2^t+t+1$ $(t\leqslant m)$ | $m$ | $2^{m-1}-2^{t-1}+2$ | 2 |
| [10], [29] | $2^{m-1}-1$ | $m$ | $2^{m-2}-1$ | 3 |
| [29] | $3\cdot 2^{m-2}$ | $m$ | $3\cdot 2^{m-3}$ | 2 |
| [33] | 45 | 30 | 4 | 8 |
| [33] | 21 | 12 | 4 | 5 |

*Construction D*. This feature of our constructions makes it possible to satisfy different design requirements on the code parameters.

2) They produce high-rate LRCs. For example, for the family of code $\mathcal{C}_{\mathrm{I}}(\mu=2)$, its code rate asymptotically approaches $\frac{r}{r+1}$ as $\ell\to\infty$. Moreover, for all of the constructed binary LRCs with $d=3$ or $d=4$, the minimum distance is optimal with respect to bound (6).

There exist several other constructions of binary $(r,1)_a$-LRCs, which are summarized in Table V. Goparaju et al. [7] and Zeh et al. [40] focused on constructing high-rate binary $(r,1)_a$-LRCs with fixed small locality 2 and small minimum distance. In [40], another construction for LRCs with arbitrary locality and fixed minimum distance 6 was given. In contrast, Silberstein et al. [29] proposed constructions of low-rate binary $(r,1)_a$-LRCs with fixed small locality but large minimum distance. In [33], Tamo et al. gave some specific examples of cyclic binary $(r,1)_a$-LRCs from subfield subcodes.

Compared to these previous code constructions, our constructions offer more flexibility with regard to the possible code parameters. First, we compare our results to those in [7], [40]. Roughly speaking, for a given length and minimum distance, our codes generally offer higher rate but at the cost of larger locality. For example, Gopparaju et al. give a $[255,162,6]$ LRC with locality $r=2$ and rate 0.6353. Zeh et al. give a $[198,155,6]$ LRC with locality $r=5$ and rate 0.7828. By comparison, referring to Table IV, we can use $\mathcal{C}_{\mathrm{I}}(\mu=3)$ and parameters $m=4$ and $\ell=16$ to construct a $[240,212,6]$ LRC with locality $r=14$ and rate 0.8833.

We also compare our constructions to some of those examples given in [33]. One example is a $[45,30,4]$ binary $(r,1)_a$-LRC with $r=8$, while we can construct a $[45,35,4]$ binary $(r,1)_a$-LRC with $r=8$ from *Construction C* using a $[13,8,4]$ binary base code $\mathcal{C}_{base}$. Another example in [33] is a $[21,12,4]$ binary $(r,1)_a$-LRC with $r=5$. In contrast, we can construct a $[20,12,4]$ binary $(r,1)_a$-LRC with $r=4$ from *Construction C* using an $[8,4,4]$ binary base code $\mathcal{C}_{base}$. In these cases, our codes offer higher rates with the same or smaller locality.

Finally, we apply bound (2) to give an upper bound on the dimension of the constructed $(r,1)_a$-LRC $\mathcal{C}_{\mathrm{I}}$, which has parameters $[n=(2^m-1)\ell, k=(2^m-2)\ell-m\sum_{i=2}^{\mu}(\lceil\frac{\mu}{i-1}\rceil-1), d=2\mu]$ and $r=2^m-2$. For $x=\ell-1$, bound (2) gives an upper bound $k_{ub}$,

$$\begin{aligned}
k_{ub} =& xr+k_{opt}^{(q)}(n-x(r+1),d)\\
=&(2^m-2)(\ell-1)+k_{opt}^{(2)}((2^m-1)\ell-(2^m-1)(\ell-1),2\mu)\\
=&(2^m-2)(\ell-1)+k_{opt}^{(2)}(2^m-1,2\mu)\\
\approx&(2^m-2)(\ell-1)+2^m-2-(\mu-1)m\\
=&(2^m-2)\ell-(\mu-1)m.
\end{aligned}$$

For small $\mu$, the gap between $k$ and the upper bound $k_{ub}$ is small, e.g., for $\mu=3$, $k_{ub}-k=m$, and for $\mu=4$, $k_{ub}-k=2m$. For large $\mu$, the gap between $k$ and $k_{ub}$ becomes large. However, it is not known whether bound (2) is tight.

## V. Binary LRCs with Availability

In this section, we study binary $(r,t)_a$-LRCs based on one-step majority-logic decodable codes [15].

**Definition 25.** *An $[n,k,d]_q$ linear code $\mathcal{C}$ is said to be a one-step majority-logic decodable code with $t$ orthogonal repair sets if the $i$th symbol, for $i\in[n]$, has $t$ pairwise disjoint repair sets $\mathcal{R}_i^j$, $j\in[t]$, such that for every $j\in[t]$ the $i$th symbol is a linear combination of all symbols in $\mathcal{R}_i^j$.*

According to Definition 25, it is evident that if $\mathcal{C}$ is a one-step majority-logic decodable code with $t$ orthogonal repair sets, and if the size of all repair sets is at most $r$, then $\mathcal{C}$ has all-symbol locality $r$ and availability $t$. Moreover, referring to a well known result (Theorem 8.1 in [15]), we can see that for an $[n,k,d]_q$ one-step majority-logic decodable code with $t$ orthogonal repair sets, all of the same size $r$, the availability $t$ satisfies

$$t\leqslant\left\lfloor\frac{n-1}{r}\right\rfloor. \tag{9}$$

Note that for a cyclic code, once $t$ repair sets are found for one symbol, the repair sets for all other symbols can be determined correspondingly from the cyclic symmetry of the code. Therefore, most of one-step majority-logic decodable codes found so far are cyclic codes. There are several constructions of one-step majority-logic decodable codes, such as doubly transitive invariant (DTI) codes, cyclic simplex codes, cyclic difference-set codes, and 4-cycle free regular linear codes [15]. The following examples present two families of one-step majority-logic decodable cyclic codes, and we give their locality and availability.

TABLE VI
DIFFERENCE-SET CODES

| $\mathcal{C}$ | $n$ | $k$ | $d$ | $r$ | $t$ | $t^u$ | $d^u$ |
|---|---|---|---|---|---|---|---|
| $m=2$ | 21 | 11 | 6 | 4 | 5 | 5 | 6 |
| $m=3$ | 73 | 45 | 10 | 8 | 9 | 9 | 12 |
| $m=4$ | 273 | 191 | 18 | 16 | 17 | 17 | 31 |
| $m=5$ | 1057 | 813 | 34 | 32 | 33 | 33 | 80 |

**Example 15.** Consider a cyclic binary simplex code with parameters $[n = 2^m - 1, k = m, d = 2^{m-1}]$. It is a one-step majority-logic decodable code with $2^{m-1} - 1$ disjoint repair sets [15]. It is easy to verify that every repair set has size 2. Therefore, it has all-symbol locality $r = 2$ and availability $t = 2^{m-1} - 1$. This code has the optimal minimum distance, due to the Plotkin bound. This locality and availability property of the simplex codes was also observed independently in [14].

**Example 16.** Consider a cyclic binary difference-set code with parameters $[n = 2^{2m} + 2^m + 1, k = 2^{2m} + 2^m - 3^m, d = 2^m + 2]$. It is a one-step majority-logic decodable code with $2^m + 1$ disjoint repair sets [15]. We can verify that every repair set has size $2^m$. Thus, this code has all-symbol locality $r = 2^m$ and availability $t = 2^m + 1$. For the codes with $2 \leqslant m \leqslant 5$, Table VI gives the upper bound $t^u$ on $t$ from bound (9) and the upper bound $d^u$ on $d$ from bound (6).

Another important class of one-step majority-logic decodable codes is 4-cycle free linear codes that have a parity-check matrix $H$ with constant row weight $\rho$ and constant column weight $\gamma$. Obviously, such codes have all-symbol locality $r = \rho - 1$ and availability $t = \gamma$. In particular, 4-cycle free $(\rho, \gamma)$-regular low-density parity-check (LDPC) codes have this property. Based upon this observation, a family of codes with all-symbol locality and availability were constructed using partial geometries in [19]. The authors of [19] also derived lower and upper bounds on the code rate; however, the exact dimension and minimum distance of these codes are still not known.

Many 4-cycle free regular LDPC codes have been constructed by leveraging different mathematical tools, e.g., finite geometries, algebraic methods, and block designs [15]. Here we consider a family of such codes based on Euclidean Geometries (EG), and we give explicit expressions for their code length, dimension, and minimum distance, as well as their locality and availability.

**Example 17.** Consider the class of binary 4-cycle free regular LDPC codes called in [15] the two-dimensional type-I cyclic $(0, m)$th-order EG-LDPC codes, with parameters $[n = 2^{2m} - 1, k = 2^{2m} - 3^m, d = 2^m + 1]$. From the structure of their parity-check matrices, they have all-symbol locality $r = 2^m - 1$ and availability $t = 2^m$. Table VII lists the parameters of these codes for $2 \leqslant m \leqslant 5$ and gives the upper bound $t^u$ on $t$ from bound (9) and the upper bound $d^u$ on $d$ from bound (6).

Finally, we briefly show how to get a long LRC with availability from a short one-step majority-logic decodable code based on a multi-level tensor product structure. We modify Step 1 in *Construction D* to provide availability

TABLE VII

TWO-DIMENSIONAL TYPE-I CYCLIC $(0, m)$TH-ORDER EG-LDPC CODES

| $\mathcal{C}$ | $n$ | $k$ | $d$ | $r$ | $t$ | $t^u$ | $d^u$ |
|---|---|---|---|---|---|---|---|
| $m = 2$ | 15 | 7 | 5 | 3 | 4 | 4 | 5 |
| $m = 3$ | 63 | 37 | 9 | 7 | 8 | 8 | 12 |
| $m = 4$ | 255 | 175 | 17 | 15 | 16 | 16 | 30 |
| $m = 5$ | 1023 | 781 | 33 | 31 | 32 | 32 | 80 |

by using the parity-check matrix of a one-step majority-logic decodable code as $H_1^{'}$. We illustrate this modification with the following example where we use for $H_1^{'}$ the parity-check matrix of the $[15, 7, 5]$ binary BCH code, which is a one-step majority-logic decodable code with all-symbol locality $r = 3$ and availability $t = 4$ [15].

**Example 18.** Let $n' = 15$ and $\alpha$ be a primitive element of $\mathbb{F}_{16}$. Let

$$H_1^{'} = \begin{bmatrix} \alpha^0 & \alpha^1 & \cdots & \alpha^{14} \\ (\alpha^3)^0 & (\alpha^3)^1 & \cdots & (\alpha^3)^{14} \end{bmatrix}$$

and $H_2^{'} = [(\alpha^5)^0, (\alpha^5)^1, \cdots, (\alpha^5)^{14}]$. Let $H_1^{''} = \mathbf{I}_{\ell\times\ell}$ and $H_2^{''} = [1, 1, \cdots, 1]$. The parity-check matrix $H$ of the constructed LRC is

$$H = \begin{bmatrix} H_1^{''} \otimes H_1^{'} \\ H_2^{''} \otimes H_2^{'} \end{bmatrix}.$$

This LRC has parameters $[15\ell, 7\ell - 2, 7]$ with all-symbol locality $r = 3$ and availability $t = 4$.

## VI. CONCLUSION

In this paper, we first studied the locality of binary classical codes and their modified versions obtained from standard code operations. We then presented several constructions of binary LRCs by using phantom parity-check symbols and a multi-level tensor product structure. Compared to other recently proposed schemes which produce binary LRCs with fixed minimum distance or locality, our constructions are more flexible and offer wider choices of the code parameters, i.e., code length, dimension, minimum distance, and locality. We also showed that our binary LRCs with minimum distance 3 or 4 are optimal with respect to the minimum distance. Finally, we studied the locality and availability properties of one-step majority-logic decodable codes, and demonstrated a construction of a long binary LRC with availability from a short one-step majority-logic decodable code.

## Appendix A
## Proof of Lemma 3

*Proof:* Assume that $x \in \mathbb{Z}^+, \boldsymbol{y} = (y_1, \ldots, y_x) \in ([t])^x$ satisfy the conditions in the lemma. Also, assume without loss of generality that the first $k$ symbols of the code $\mathcal{C}$ form an information set.

The set $\mathcal{I}$ is constructed according to the following procedure.

**Procedure A**

1) Let $\mathcal{I}_0 = \emptyset$.
2) **For** $j = 1, \ldots, x$
3) Choose an integer $a_j \in [k]$ and $a_j \notin \mathcal{I}_{j-1}$, such that
   $k_{\mathcal{I}_{j-1} \cup \{a_j\}} = k_{\mathcal{I}_{j-1}} + 1$.
4) $\mathcal{I}_j = \mathcal{I}_{j-1} \cup \{a_j\} \cup \mathcal{R}^1_{a_j} \cup \cdots \cup \mathcal{R}^{y_j}_{a_j}$.
5) **End**
6) Let $\mathcal{I} = \mathcal{I}_x \cup \mathcal{S}$, where $\mathcal{S} \subseteq [n] \setminus \mathcal{I}_x$ is a set of cardinality $\min\{n, B(r, x, \boldsymbol{y})\} - |\mathcal{I}_x|$. □

This completes the construction of the set $\mathcal{I}$.

First, let us show that the construction of the set $\mathcal{I}$ is well defined.

**Claim 1** *In step 3), it is always possible to find a coordinate $a_j \in [k]$, for $1 \leqslant j \leqslant x$, that satisfies the condition in this step.*

*Proof:* To see this, we show that on the $j$th loop, for $1 \leqslant j \leqslant x$, the value of $k_{\mathcal{I}_{j-1}}$ satisfies $k_{\mathcal{I}_{j-1}} < k$, and thus at least one of the first $k$ coordinates does not belong to the set $\mathcal{I}_{j-1}$. Since the value of $k_{\mathcal{I}_{j-1}}$ increases with $j$, it is enough to show that $k_{\mathcal{I}_{x-1}} \leqslant k - 1$.

Let $\mathcal{S}_{a_j} = \{a_j\} \cup \mathcal{R}^1_{a_j} \cup \cdots \cup \mathcal{R}^{y_j}_{a_j}$ for $j \in [x]$. First, we show that $k_{\mathcal{S}_{a_j}} \leqslant (r-1)t + 1$. Let $G = [g_1, \ldots, g_n]$ be a generator matrix of the code $\mathcal{C}$. For the repair set $\mathcal{R}^u_{a_j}$, $u \in [y_j]$, $g_{a_j}$ is a linear combination of the columns $g_m$, $m \in \mathcal{R}^u_{a_j}$, so there exists a coordinate $b^u_j \in \mathcal{R}^u_{a_j}$ such that $g_{a_j} = \sum_{m \in \mathcal{R}^u_{a_j} \setminus \{b^u_j\}} \alpha_m g_m + \beta_{b^u_j} g_{b^u_j}$, where $\alpha_m, \beta_{b^u_j} \in \mathbb{F}_q$ and $\beta_{b^u_j} \neq 0$. Thus, $k_{\{a_j\} \cup \mathcal{R}^u_{a_j} \setminus \{b^u_j\}} = k_{\{a_j\} \cup \mathcal{R}^u_{a_j}}$. Therefore, we have

$$\begin{aligned} k_{\mathcal{S}_{a_j}} =& k_{\mathcal{S}_{a_j} \setminus \{\cup_{u=1}^{y_j} b^u_j\}} \overset{(a)}{\leqslant} |\mathcal{S}_{a_j} \setminus \{\cup_{u=1}^{y_j} b^u_j\}| \leqslant (r-1)y_j + 1 \\ \leqslant& (r-1)t + 1, \end{aligned}$$

where (a) follows from the fact that $k_{\mathcal{M}} \leqslant |\mathcal{M}|$ for any set $\mathcal{M} \subseteq [n]$.

From the construction of the set $\mathcal{I}$, we have that $\mathcal{I}_{x-1} = \cup_{j=1}^{x-1}\mathcal{S}_{a_j}$ and therefore

$$\begin{aligned} k_{\mathcal{I}_{x-1}} = k_{\cup_{j=1}^{x-1}\mathcal{S}_{a_j}} &\overset{(a)}{\leqslant} \sum_{j=1}^{x-1} k_{\mathcal{S}_{a_j}} \\ &\overset{(b)}{\leqslant} (x-1)[(r-1)t+1] \\ &\overset{(c)}{\leqslant} \left(\left\lceil \frac{k-1}{(r-1)t+1} \right\rceil - 1\right)[(r-1)t+1] \\ &< \frac{k-1}{(r-1)t+1}[(r-1)t+1] = k-1, \end{aligned}$$

where $(a)$ follows from the fact that $k_{\mathcal{M}_1 \cup \mathcal{M}_2} \leqslant k_{\mathcal{M}_1} + k_{\mathcal{M}_2}$ for any sets $\mathcal{M}_1, \mathcal{M}_2 \subseteq [n]$ and a simple induction. Inequality $(b)$ follows from $k_{\mathcal{S}_{a_j}} \leqslant (r-1)t+1$, and $(c)$ follows from $x \leqslant \lceil \frac{k-1}{(r-1)t+1} \rceil$. ■

It is clear to see that the set $\mathcal{I}$ has size of $|\mathcal{I}| = \min\{n, B(r,x,\boldsymbol{y})\}$.

Next, we show that $k_{\mathcal{I}} \leqslant A(r,x,\boldsymbol{y})$. To do this, in **Procedure A**, for each $j$th iteration, let us add the following coordinate selection steps between step 3) and step 4).

3.1) **For** $\ell = 1, \ldots, y_j$

3.2) Choose an integer $a_j^\ell \in \mathcal{R}_{a_j}^\ell$ and $a_j^\ell \notin \mathcal{I}_{j-1}$, such that $k_{\{a_j\}\cup\mathcal{R}_{a_j}^\ell \backslash \{a_j^\ell\}} = k_{\{a_j\}\cup\mathcal{R}_{a_j}^\ell}$.

3.3) **End**

We next show that the above steps are well defined.

**Claim 2** *In step 3.2), it is always possible to find an integer $a_j^\ell$, for $1 \leqslant j \leqslant x$ and $1 \leqslant \ell \leqslant y_j$, that satisfies the condition in this step.*

*Proof:* First, assume on the contrary that $\mathcal{R}_{a_j}^\ell \subseteq \mathcal{I}_{j-1}$. Then, we conclude that $k_{\mathcal{I}_{j-1}\cup\{a_j\}} = k_{\mathcal{I}_{j-1}}$, which violates the selection rule in step 3). Second, for the case of $\mathcal{R}_{a_j}^\ell \nsubseteq \mathcal{I}_{j-1}$, since $g_{a_j}$ is a linear combination of $g_i$, $i \in \mathcal{R}_{a_j}^\ell$, there exists at least one coordinate $a_j^\ell \in \mathcal{R}_{a_j}^\ell$ and $a_j^\ell \notin \mathcal{I}_{j-1}$ such that $g_{a_j} = \sum_{i\in\mathcal{R}_{a_j}^\ell\backslash\{a_j^\ell\}} \alpha_i g_i + \beta_{a_j^\ell} g_{a_j^\ell}$, where $\alpha_i, \beta_{a_j^\ell} \in \mathbb{F}_q$ and $\beta_{a_j^\ell} \neq 0$. Therefore, $g_{a_j^\ell}$ can be expressed as a linear combination of the columns $g_i$, for $i \in \{a_j\} \cup \mathcal{R}_{a_j}^\ell \backslash \{a_j^\ell\}$, so we have $k_{\{a_j\}\cup\mathcal{R}_{a_j}^\ell\backslash\{a_j^\ell\}} = k_{\{a_j\}\cup\mathcal{R}_{a_j}^\ell}$. ■

Now, let $\mathcal{P}$ be the set of coordinates chosen in steps 3.1) – 3.3): $\mathcal{P} = \{a_1^1, \ldots, a_1^{y_1}, \ldots, a_x^1, \ldots, a_x^{y_x}\}$. From the construction, the integers $a_1^1, \ldots, a_1^{y_1}, \ldots, a_x^1, \ldots, a_x^{y_x}$ are all different, i.e., $|\mathcal{P}| = \sum_{j=1}^{x} y_j$.

Next, we prove that $k_{\mathcal{I}} \leqslant A(r,x,\boldsymbol{y})$ by showing that $k_{\mathcal{I}\backslash\mathcal{P}} \leqslant A(r,x,\boldsymbol{y})$ and $k_{\mathcal{I}} = k_{\mathcal{I}\backslash\mathcal{P}}$.

**Claim 3** *$k_{\mathcal{I}\backslash\mathcal{P}} \leqslant A(r,x,\boldsymbol{y})$.*

*Proof:*

$$\begin{aligned} k_{\mathcal{I}\backslash\mathcal{P}} &\leqslant |\mathcal{I}\backslash\mathcal{P}| \overset{(a)}{=} \min\{n, B(r,x,\boldsymbol{y})\} - \sum_{j=1}^{x} y_j \\ &\leqslant B(r,x,\boldsymbol{y}) - \sum_{j=1}^{x} y_j = A(r,x,\boldsymbol{y}), \end{aligned}$$

where $(a)$ follows from $|\mathcal{I}| = \min\{n, B(r,x,\boldsymbol{y})\}$ and $|\mathcal{P}| = \sum_{j=1}^{x} y_j$. ■

**Claim 4** $k_{\mathcal{I}} = k_{\mathcal{I}\setminus\mathcal{P}}$.

*Proof:* Showing that $k_{\mathcal{I}} = k_{\mathcal{I}\setminus\mathcal{P}}$ is equivalent to showing that for any two codewords $\boldsymbol{c}$ and $\hat{\boldsymbol{c}}$ in code $\mathcal{C}$, if $\boldsymbol{c}_{\mathcal{I}\setminus\mathcal{P}} = \hat{\boldsymbol{c}}_{\mathcal{I}\setminus\mathcal{P}}$, then $\boldsymbol{c}_{\mathcal{P}} = \hat{\boldsymbol{c}}_{\mathcal{P}}$.

Assume on the contrary that there exist two codewords $\boldsymbol{c} = (c_1,\ldots,c_n)$ and $\hat{\boldsymbol{c}} = (\hat{c}_1,\ldots,\hat{c}_n)$ in code $\mathcal{C}$ that $\boldsymbol{c}_{\mathcal{I}\setminus\mathcal{P}} = \hat{\boldsymbol{c}}_{\mathcal{I}\setminus\mathcal{P}}$, but $\boldsymbol{c}_{\mathcal{P}} \neq \hat{\boldsymbol{c}}_{\mathcal{P}}$. Let $\mathcal{E} = \{i : c_i \neq \hat{c}_i,\ i \in \mathcal{P}\}$. We order the elements in $\mathcal{E}$ according to the lexicographical order $\prec$ defined as follows:

1. If $i < j$, then $a_i^u \prec a_j^v$, for $i,j \in [x]$, $u \in [y_i]$, and $v \in [y_j]$.
2. If $u < v$, then $a_i^u \prec a_i^v$, for $i \in [x]$, $u,v \in [y_i]$.

Suppose that the smallest element with respect to the lexicographical order $\prec$ in $\mathcal{E}$ is $a_i^u$. According to the construction steps, we have $(\{a_i\} \cup \mathcal{R}_{a_i}^u \setminus \{a_i^u\}) \bigcap \mathcal{E} = \emptyset$ and $(\{a_i\} \cup \mathcal{R}_{a_i}^u \setminus \{a_i^u\}) \subseteq \mathcal{I}$. Since $\boldsymbol{c}_{\mathcal{I}\setminus\mathcal{E}} = \hat{\boldsymbol{c}}_{\mathcal{I}\setminus\mathcal{E}}$, we have $\boldsymbol{c}_{\{a_i\}\cup\mathcal{R}_{a_i}^u\setminus\{a_i^u\}} = \hat{\boldsymbol{c}}_{\{a_i\}\cup\mathcal{R}_{a_i}^u\setminus\{a_i^u\}}$, but $c_{a_i^u} \neq \hat{c}_{a_i^u}$. This violates the selection rule in step 3.2) for $a_i^u$: $k_{\{a_i\}\cup\mathcal{R}_{a_i}^u\setminus\{a_i^u\}} = k_{\{a_i\}\cup\mathcal{R}_{a_i}^u}$, which indicates that if $\boldsymbol{c}_{\{a_i\}\cup\mathcal{R}_{a_i}^u\setminus\{a_i^u\}} = \hat{\boldsymbol{c}}_{\{a_i\}\cup\mathcal{R}_{a_i}^u\setminus\{a_i^u\}}$ then $c_{a_i^u} = \hat{c}_{a_i^u}$. Thus, we get a contradiction and conclude that there do not exist two codewords $\boldsymbol{c}$ and $\hat{\boldsymbol{c}}$ in code $\mathcal{C}$ that $\boldsymbol{c}_{\mathcal{I}\setminus\mathcal{P}} = \hat{\boldsymbol{c}}_{\mathcal{I}\setminus\mathcal{P}}$, but $\boldsymbol{c}_{\mathcal{P}} \neq \hat{\boldsymbol{c}}_{\mathcal{P}}$. ■

From Claims 3 and 4, it is clear to see that we have $k_{\mathcal{I}} \leqslant A(r,x,\boldsymbol{y})$. Therefore, there exists a set $\mathcal{I} \subseteq [n]$, $|\mathcal{I}| = \min\{n, B(r,x,\boldsymbol{y})\}$, such that $k_{\mathcal{I}} \leqslant A(r,x,\boldsymbol{y})$. Finally, since $k_{\mathcal{I}} \leqslant A(r,x,\boldsymbol{y}) < k$ and $k_{[n]} = k$, we conclude that $B(r,x,\boldsymbol{y}) < n$ and $|\mathcal{I}| = \min\{n, B(r,x,\boldsymbol{y})\} = B(r,x,\boldsymbol{y})$.

■

## APPENDIX B
## PROOF OF THEOREM 4

*Proof:* We follow similar steps to the proof in [4] which consists of two parts. First, from Lemma 3, for any $[n,k,d]_q$ linear code $\mathcal{C}$ with information locality $r$ and availability $t$, for all $x \in \mathbb{Z}^+$ and $\boldsymbol{y} = (y_1,\ldots,y_x) \in ([t])^x$ satisfying $1 \leqslant x \leqslant \lceil \frac{k-1}{(r-1)t+1} \rceil$ and $A(r,x,\boldsymbol{y}) < k$, there exists a set $\mathcal{I} \subseteq [n]$, $|\mathcal{I}| = B(r,x,\boldsymbol{y})$, such that $k_{\mathcal{I}} \leqslant A(r,x,\boldsymbol{y})$.

For the second part of the proof, for any $x \in \mathbb{Z}^+$ and $\boldsymbol{y} = (y_1,\ldots,y_x) \in ([t])^x$, the $\mathcal{I} \subseteq [n]$ is constructed as in the first part. Then, we consider the code $\mathcal{C}_{\mathcal{I}}^{\boldsymbol{0}} = \{\boldsymbol{c}_{[n]\setminus\mathcal{I}} : \boldsymbol{c}_{\mathcal{I}} = \boldsymbol{0} \text{ and } \boldsymbol{c} \in \mathcal{C}\}$. Since the code $\mathcal{C}$ is linear, the size of the code $\mathcal{C}_{\mathcal{I}}^{\boldsymbol{0}}$ is $q^{k-k_{\mathcal{I}}}$ and it is a linear code as well. Moreover, the minimum distance $D$ of the code $\mathcal{C}_{\mathcal{I}}^{\boldsymbol{0}}$ is at least $d$, i.e., $D \geqslant d$.

Thus, we get an upper bound on the minimum distance $d$,

$$\begin{aligned} d \leqslant D \leqslant & d_{\ell-opt}^{(q)}[n - |\mathcal{I}|, k - k_{\mathcal{I}}] \\ \leqslant & d_{\ell-opt}^{(q)}[n - |\mathcal{I}|, k - A(r,x,\boldsymbol{y})]. \end{aligned}$$

Therefore, we conclude that

$$\begin{aligned} d \leqslant & d_{\ell-opt}^{(q)}[n - |\mathcal{I}|, k - A(r,x,\boldsymbol{y})] \\ = & d_{\ell-opt}^{(q)}[n - B(r,x,\boldsymbol{y}), k - A(r,x,\boldsymbol{y})]. \end{aligned}$$

Similarly, we also get an upper bound on the dimension $k$,

$$k - k_{\mathcal{I}} \leqslant k_{\ell-opt}^{(q)}[n - |\mathcal{I}|, D] \leqslant k_{\ell-opt}^{(q)}[n - |\mathcal{I}|, d].$$

Therefore, we conclude that

$$\begin{aligned} k \leqslant & k_{\ell-opt}^{(q)}[n - |\mathcal{I}|, d] + k_{\mathcal{I}} \\ \leqslant & k_{\ell-opt}^{(q)}[n - B(r, x, \boldsymbol{y}), d] + A(r, x, \boldsymbol{y}). \end{aligned}$$

■

## APPENDIX C
## PROOF OF LEMMA 13

*Proof:* We prove that the minimum distance of the constructed $(r,1)_i$-LRC from *Construction A* is at least 3 by verifying that it can correct any two erasures. We consider the following 2-erasure patterns, where we refer to their locations in the blocks in Fig. 1. We refer to block **I-II** as the union of block **I** and block **II**.

1) Two erasures are in the same row in block **I-II**, e.g., $\mu_{11}$ and $p_{L_1}$ are erased in the first row. We can first recover parity-check symbols $(p_{11}, \ldots, p_{1,n'-k'})$, based on which two erased symbols can be recovered.

2) Two erasures in different rows in block **I-II** can be recovered individually from the local parity-check equation.

3) Two erasures in block **IV** can be recovered from all existing information symbols.

4) One erasure is in block **I-II** and one erasure is in block **IV**. First, the erasure in block **I-II** can be recovered from the local parity-check equation. Then, the erasure in block **IV** can be recovered from all the existing information symbols.

The proof for the constructed $(r,1)_i$-LRC from *Construction A'* follows the same ideas and is thus omitted.

■

## APPENDIX D
## PROOF OF THEOREM 20

*Proof:* From *Construction B*, the code length, dimension, and locality are determined. As in the proof of Lemma 13, to prove that the minimum distance of the $(r,1)_i$-LRC is at least 5, we only need to enumerate all possible 4-erasure patterns and then verify they can be corrected. In the following, we show how to recover two typical 4-erasure patterns in Fig. 2. Other patterns can be verified in a similar way and hence are omitted.

1) There are two erasures in a row in block **I** and two erasures in another row in block **I**. Without loss of generality, assume they appear on the first two rows $(\mu_{11}, \ldots, \mu_{1k'})$ and $(\mu_{21}, \ldots, \mu_{2k'})$ in block **I**. We can recover this 4-erasure pattern as follows. First, with $(p_{G_{11}}, \ldots, p_{G_{1,w}})$ and $(p_{G_{21}}, \ldots, p_{G_{2,w}})$, we can recover $(p_{11}, \ldots, p_{1,w})$ and $(p_{21}, \ldots, p_{2,w})$. Then, two erasures in the first row $(\mu_{11}, \ldots, \mu_{1k'})$ can be recovered since only two erasures appear in the codeword $(\mu_{11}, \ldots, \mu_{1k'}, p_{11}, \ldots, p_{1,w})$ which belongs to a code with minimum distance at least 3. Similarly, two erasures in the second row $(\mu_{21}, \ldots, \mu_{2k'})$ can be recovered since only two erasures appear in the codeword $(\mu_{21}, \ldots, \mu_{2k'}, p_{21}, \ldots, p_{2,w})$.

2) There are two erasures in a row in block **I**, one erasure in $(p_{G_{11}}, \ldots, p_{G_{1,w}})$ in block **IV**, and one more erasure in $(p_{G_{21}}, \ldots, p_{G_{2,w}})$ in block **IV**. For simplicity, we assume that the erasures are located in positions $\mu_{11}$, $\mu_{12}$, $p_{G_{11}}$, and $p_{G_{21}}$. We can recover this 4-erasure pattern as follows. First, with $(p_{G_{12}}, \ldots, p_{G_{1,n'-k'}})$, we can recover $(p_{12}, \ldots, p_{1,n'-k'})$. Then, $\mu_{11}$ and $\mu_{12}$ can be recovered since only three erasures appear in the codeword $(\mu_{11}, \ldots, \mu_{1k'}, p_{11}, \ldots, p_{1,n'-k'})$. Finally, $p_{G_{11}}$ and $p_{G_{21}}$ can be recovered.

■